\documentclass[pra,aps,twocolumn,nopacs,superscriptaddress,nofootinbib]{revtex4-1}

\usepackage{amsmath}  \usepackage{amssymb}   \usepackage{amsfonts}  \usepackage{bm}  \usepackage{bbm}      \usepackage{braket}    \usepackage{color}  \usepackage{comment}  \usepackage{dcolumn}  \usepackage{enumerate}  \usepackage{epsfig}  \usepackage{gensymb}  \usepackage{graphicx}  \usepackage{indentfirst}  \usepackage{lmodern}  \usepackage{mathrsfs}  \usepackage{mathtools}  \usepackage{psfrag}   \usepackage{pst-all}   \usepackage{soul}  
\usepackage{xcolor}
\usepackage{upgreek}
\usepackage{float} 
\usepackage[colorlinks,linkcolor=blue,citecolor=blue,urlcolor=blue,hyperindex,driverfallback=dvipdfm]{hyperref}  \usepackage[T1]{fontenc} 

\def\ii{{\rm i}}  \def\ee{{\rm e}}
\def\me{m_{\rm e}}  
\def\Ree{{\rm Re}}  \def\Imm{{\rm Im}}
\def\Ab{{\bf A}}        \def\Eb{{\bf E}}                  \def\jb{{\bf j}}              \def\Qb{{\bf Q}}    \def\Rb{{\bf R}}  \def\rb{{\bf r}}       \def\Db{{\bf D}}  \def\nb{{\bf n}}  \def\ssb{{\bf s}}
                
      \def\wp{\omega_{\rm p}}   
      
\def\Qc{Q_\mathrm{c}}  \def\Qf{Q_\mathrm{f}}
\def\lf{\ell_\mathrm{f}} \def\li{\ell_\mathrm{i}}
\def\der{\mathrm{d}} \def\Rcut{R_\mathrm{c}} \def\alphaf{\alpha_\mathrm{f}}
\def\BW{\mathrm{s}_0}
\allowdisplaybreaks
\begin{document} 
\def\bibsection{\section*{\refname}} 

\title{Optimal conditions for detecting optical dichroism at the nanoscale by electron energy-loss spectroscopy
}


\author{Marek Z\'{a}le\v{s}\'{a}k}
\affiliation{Brno University of Technology, Faculty of Mechanical Engineering, Institute of Physical Engineering, Technická 2, 61669 Brno, Czech Republic}
\affiliation{Brno University of Technology, Central European Institute of Technology, Purkyňova 123, 61200 Brno, Czech Republic}

\author{Martin O\v{s}mera}
\affiliation{Brno University of Technology, Faculty of Mechanical Engineering, Institute of Physical Engineering, Technická 2, 61669 Brno, Czech Republic}
\affiliation{Uppsala University, Faculty of Science and Technology, Department of Physics and Astronomy, Ångströmlaboratoriet, Regementsvägen 10, 752 37 Uppsala, Sweden}

\author{Martin Hrto\v{n}}
\affiliation{Brno University of Technology, Faculty of Mechanical Engineering, Institute of Physical Engineering, Technická 2, 61669 Brno, Czech Republic}
\affiliation{Brno University of Technology, Central European Institute of Technology, Purkyňova 123, 61200 Brno, Czech Republic}

\author{Andrea Kone\v{c}n\'{a}}
\email{andrea.konecna@vutbr.cz}
\affiliation{Brno University of Technology, Faculty of Mechanical Engineering, Institute of Physical Engineering, Technická 2, 61669 Brno, Czech Republic}
\affiliation{Brno University of Technology, Central European Institute of Technology, Purkyňova 123, 61200 Brno, Czech Republic}

\begin{abstract}
The emergence of optical circular dichroism in chiral nanoscale and molecular systems provides not only a way for analyzing the sample chirality itself but also additional degrees of freedom in manipulating light. Such manipulation can be reached even at the nanoscale level; however, probing and understanding the properties of optical fields well below the diffraction limit requires an adequate technique. Electron energy-loss spectroscopy (EELS) with orbital angular momentum (OAM)-based electron state sorting has been suggested as a suitable candidate, but to date, no conclusive experiments have been performed. We, therefore, theoretically explore the emergence of dichroism in EELS for a canonical single-twist helix nanostructure and present a detailed analysis of the optimal parameters to obtain a robust signal. Our work offers novel insights into the interpretation and volatility of the OAM-resolved EELS signal, which can inspire and guide future experimental efforts.
\end{abstract}

\maketitle
\date{\today}


Chirality, the lack of mirror symmetry, is present in many naturally occurring nanoscale systems, such as in amino acids or protein molecules, and it plays an important role in various biological and chemical processes \cite{guijarro2008originofchirality}. The chirality can be revealed from direct structural measurements, but it also manifests itself in vibrational or electronic properties, which results in a different response of a chiral molecule to circularly polarized light -- circular dichroism \cite{rodger1997circular,magyarfalvi2011vibrational}. The difference in the absorption or emission of light by different enantiomers is further utilized in the design of artificially fabricated chiral nanostructures \cite{fan_chiral_2012,yin_interpreting_2013,schaferling2017chiral,Kwon2023_chiralSpectroscopyofNanostructures}, which find applications, e.g., in light manipulation \cite{Gansel2009_polarizer,schreiber2013chiral,jen2017densely, huang_chiroptically_2017,Collins2018_chiralMetasurface,Karakasoglu2018_polarizationControl} or in the so-called enhanced spectroscopies relying on the enhancement of the dichroic signal of chiral molecules placed in the tailored optical near field of a chiral nanostructure \cite{Warning2021_chiralitySensing, adhikari_optically_2022}.














Although light represents a relatively non-destructive and effective way to probe even sensitive chiral molecules \cite{penasa_advances_2025}, light-based techniques are diffraction-limited. Probing the dichroic response from a small number of molecules \cite{hassey2006_dichroism_molecule} or even spatial mapping of the dichroic response at sub-wavelength spatial resolution, which could be particularly useful for understanding the artificially designed chiral near fields \cite{Schaferling2012_TailoringEnhancedOpticalChirality,Schaferling2014_HelicalPlasmonicNanoparticles}, thus remains challenging. As an alternative to light, an electron beam represents a suitable probe that can provide spatial resolution down to the single-atom level and simultaneously measure the low-energy vibrational and electronic response of matter \cite{GDA,krivanek_vibrational_2014,Yang2022,Roadmap2025}.

It has been theoretically predicted that information on optical and vibrational chirality can be retrieved in electron energy-loss spectroscopy (EELS) experiments that allow for the detection of electrons based on the orbital angular momentum (OAM) transferred in their interaction with the sample \cite{Asenjo_garcia,Zanfrognini2019,lourenco_NatPhys,Guido2021,Bourgeois2023,Garrigou2025}. Such OAM-based sorting can now be achieved with electron sorters that rely on electrostatic fields \cite{Grillo2017, McMorran2017, Tavabi2021}. One can also consider the possibility of a suitably designed initial electron state that could further enhance the capability of detecting the desired dichroic signal, e.g., a vortex electron beam \cite{Bliokh2017,Lloyd2017}, which can be generated using approaches based on diffraction \cite{Verbeeck_2010,MAA11,mafakheri_realization_2017}, sculpted thin films \cite{Uchida2010}, or the interaction with static and dynamic electromagnetic fields \cite{Beche2014,Pozzi2017, Tavabi2020, Tavabi2022,Vanacore2019,Kozak2021, Yu2023_48progphaseplate}.

However, despite the promising theoretical predictions, the experimental detection of nanoscale dichroism due to low-energy excitations in EELS has been inconclusive so far \cite{harvey}, which is even more surprising given the increasing capabilities in generating and detecting states with OAM \cite{Tavabi2021,Tavabi2022}. In this Letter, we therefore develop a theoretical formalism based on a decomposition of the sample potential in a suitable basis, providing insights into the role of experimental parameters on the resulting dichroic signal. We apply our theoretical findings on a prototypical chiral nanostructure -- a plasmonic nanohelix -- and demonstrate that dichroism in EELS can exhibit both sign flips and differ over orders of magnitude based on the probing geometry and electron microscope settings. 



\begin{figure*}
    \centering
    \includegraphics[width=.8\linewidth]{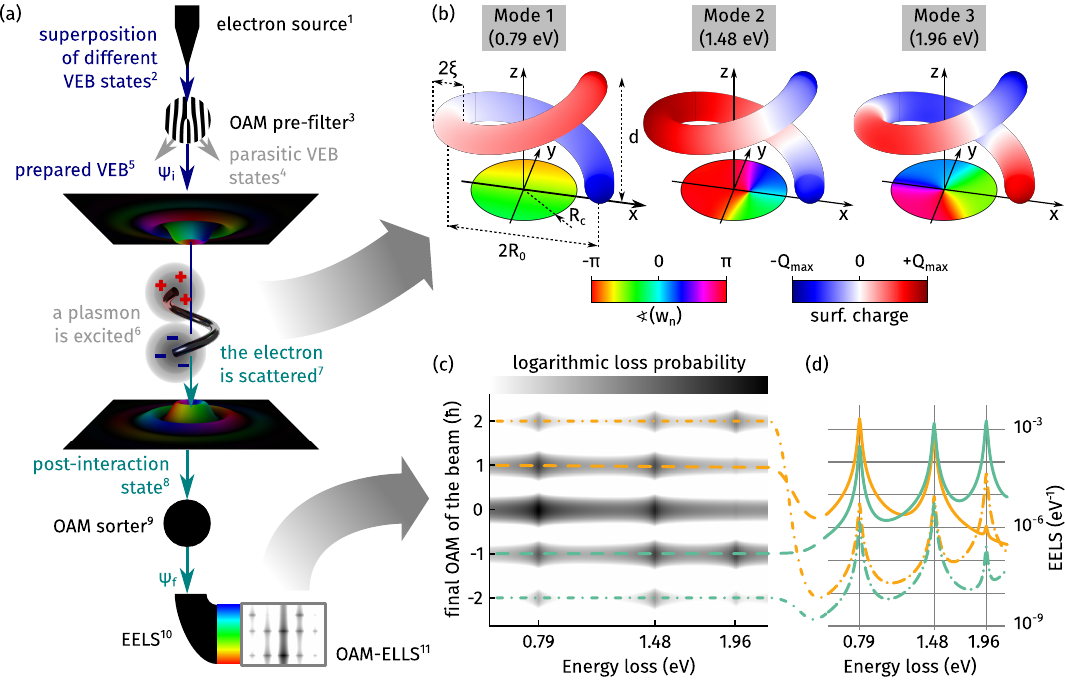}
    \caption{Overview of proposed experiment: (a) A scheme of the setup. Electrons from the source (1) in the superposition of different VEB states (2) pass through an OAM pre-filter (3). A VEB with a well-defined OAM is prepared (5) and the states with different-than-chosen OAM are sorted out (4). A VEB passing near the sample can excite a plasmon (6) and scatter (7). After the interaction, the beam is in a statistically mixed state (8). An OAM sorter (9) is used to separate the mixed states. The electrons are further dispersed in energy by the EELS spectrometer (10).
    (b) Pseudo-charge density and phase of the projected eigenpotentials $w_n$ corresponding to the first three modes of a silver helix with thickness $2 \xi=20\,\mathrm{nm}$, pitch $d=100\,\mathrm{nm}$, and radius $R_0=50\,\mathrm{nm}$.  
    (c) Calculated OAM-resolved EELS image for the sample from (b) illuminated by a non-vortex Gaussian beam ($\li=0$, $\BW=10$~nm, $\alphaf=5$~mrad). The image is smeared in the (vertical) OAM axis by a Gaussian to illustrate the finite resolution of the OAM sorter. (d) A cut-through for direct comparison between spectra obtained for final OAMs $\pm \hbar$ and $\pm 2\hbar$, respectively.
    } 
    \label{fig:Fig1}
\end{figure*}  

In the following, we consider a hypothetical experimental setup sketched in Fig.~\ref{fig:Fig1}(a). We assume that with a suitable method (e.g., using diffraction-based phase plates), we obtain a well-defined initial beam characterized by $\psi_\mathrm{i}(R,\phi)=\exp(\ii  q_z  z)\exp(\ii  \li  \phi)f_{\li}(R)$, where $\hbar \, q_z$ is the component of the electron momentum along the beam axis (with the reduced Planck constant $\hbar$), $f_{\li}(R)$ is a function dependent on the transverse distance from the origin $R=\sqrt{x^2+y^2}$, and $\phi$ is the azimuthal angle. The electrons can have either zero or non-zero topological charge $\li$, which quantizes the OAM $L_z = \hbar\li$. The incident electrons then interact with the sample, and some of them exchange energy or even OAM through the optical field. The energy loss and OAM transfer can be analyzed afterward using an OAM analyzer and an EELS spectrometer. For instance, the combination of an OAM sorter and an EELS spectrometer \cite{tavabi2025demonstration} could directly yield energy dispersion and OAM dispersion (due to the finite resolution of the OAM sorter) along the two perpendicular axes of the EELS camera as illustrated in Fig.~\ref{fig:Fig1}(c).

The electron energy-loss probability corresponding to a final state with a specific final OAM $\hbar\lf$ can be, in the non-retarded and non-recoil approximation, expressed as \cite{Zanfrognini2019}
\begin{align}
    \Gamma_{\li}^{\lf}(\omega)&=\frac{e^2 }{2 \pi^2 \hbar \, v^2}\sum_{n=1}^{\infty} g_n(\omega)\int_0^{Q_\mathrm{c,f}} \Qf \, \mathrm{d} \Qf\,\left\lvert\int R \, \mathrm{d} R\right . \nonumber\\ 
    &\left.\times\int_0^{2\pi}\mathrm{d}\phi\,f_{\li}(R)\, \ee^{\ii\Delta\ell\phi}J_{\lf}(\Qf \, R) \, w_n(R,\phi,\omega_{n})\right\rvert^2,
    \label{Eq:Gamma_lfNR}
\end{align}
where $e$ is the elementary charge, $v$ is the relativistic electron velocity, $\Delta \ell=\lf-\li$ is the topological charge transfer, and we considered that a detector centered at $R=0$ collects electrons with transverse wavevectors $\Qf$ up to a cutoff $Q_\mathrm{c,f}=\mathrm{sin}(\alphaf)\sqrt{2\me\hbar\epsilon_\mathrm{i}(1+\hbar\epsilon_\mathrm{i}/(2\me\, c^2))}/\hbar$ corresponding to the collection angle $\alphaf$. $\me$ is the electron mass, $\hbar\epsilon_\mathrm{i}$ is the initial electron energy, and $c$ is the speed of light in vacuum. $J_{\lf}(x)$ are the Bessel functions of the first kind and order $\lf$. Properties of the sample excitations are embedded in the functions $g_n$ and $w_n$, which introduce spectral and spatial dependence, respectively, of the nanostructure's optical eigenmodes indexed by $n$ [see Sec.~C in Supplemental Material (SM) \cite{SM}]. These functions are related to the screened potential $W(\rb,\rb',\omega)$ as 
$\int\mathrm{d}z\int\mathrm{d}z'\,\mathrm{Im}\left\lbrace - W(\rb,\rb',\omega)\right\rbrace\ee^{-\ii\omega(z-z')/v}
\approx \sum_{n=1}^\infty g_n(\omega) \, w_n(\Rb,\omega_n) \, w_n^\ast(\Rb',\omega_{n})$,
where we considered well-defined, spectrally not-overlapping modes resonating at frequencies $\omega_n$. 

We can immediately notice from Eq.~\eqref{Eq:Gamma_lfNR} that to ensure high spectral magnitudes necessary to overcome noise in EELS experiments, one has to optimise the spatial overlap of the projected sample potential in terms of $w_n$ with the initial and final transverse wave function profiles. As the optical near fields are typically extended over tens of nanometers, it is therefore desirable to work with not-so-well-focused beams (previously considered in Refs.~\cite{Mohammadi,Konecna2023}). We now apply Eq.~\eqref{Eq:Gamma_lfNR} for the transition from the initial non-vortex state (OAM $\hbar\li=0$) to the selected final state with OAM $\hbar\, \lf$ in the interaction with a silver single-twist nanohelix (see Fig.~\ref{fig:Fig1}(b) for the geometry and selected parameters). The nanohelix is a suitable candidate for demonstrating our formalism because it features clear dichroism when probed by circularly polarized light \cite{Hoflich2019_singleHelix_resonant_behavior, Wozniak_Chiroptical_2018}. 
Moreover, nanohelices can be relatively routinely fabricated in a variety of dimensions using different types of deposition techniques \cite{Caridad2014_helixNanofabrication,Esposito2015,Kosters2017_CoreShellNanohelices,Reisecker2024}. Eigenmodes and the corresponding eigenpotentials of the nanohelix plotted in Fig.~\ref{fig:Fig1}(b) are calculated in MNPBEM \cite{Hohenester2012}, where we also verified validity of the non-retarded approximation and spectral separation of the modes with silver as the helix's material (for details and the optical response of silver, see  Sec.~C in SM \cite{SM}). We consider a Gaussian transverse profile of the initial wave function with $f_0(R)=\sqrt{2/(\pi \BW^2)}\, \mathrm{exp}(-R^2/\BW^2)$, where we consider the beam waist $\BW=10$~nm to minimize interaction of the beam with the helix's bulk. The calculation is performed for the perfectly aligned axis of the electron beam and nanohelix, for an electron microscope operated at the acceleration voltage of 30 kV (electron velocity $v=0.328\, c$) with a collection angle $\alphaf=5$~mrad, which are feasible parameters for EELS in a scanning transmission electron microscope. Fig.~\ref{fig:Fig1}(c) summarizes the spectra calculated for different $\lf$'s that could be distinguished using the OAM sorter. The selected spectra obtained for $\Delta\ell=+1, +2$ and $\Delta\ell=-1, -2$ plotted separately in Fig.~\ref{fig:Fig1}(d) exhibit clear differences and therefore demonstrate the emergence of dichroism in EELS. Importantly, we can notice that the magnitude of the EELS signal rapidly decreases with increasing $\Delta\ell$ [note the logarithmic scale in Fig.~\ref{fig:Fig1}(c,d)], which already indicates that high-OAM transitions might not be favorable when probing optical dichroism.

To further understand the results presented in Fig.~\ref{fig:Fig1} and potentially improve the efficiency and robustness of the dichroic signal, it is beneficial to develop an analytical model. We employ the decomposition of the spatial eigenpotential dependence $w_n$ in the Fourier-Bessel basis, which is particularly suitable due to the cylindrical symmetry of the system:
\begin{align}
w_n(R,\phi,\omega_{n})=\sum_{m=-\infty}^\infty\ii^m\ee^{\ii m\phi}\sum_{j=1}^\infty J_m\left(\alpha_{j,m}\frac{R}{a}\right)A_{j,m}^n,
\label{Eq:un}
\end{align}
where $m$ are integers, $a$ is a finite radius of an area, where we evaluate the projected potential, and $A_{j,m}^n$ are expansion coefficients expressed as $A_{j,m}^n=\frac{(-\ii)^m}{\pi a^2 \left[J_{m+1}\left(\alpha_{j,m}\right)\right]^2 }\int_0^a R \, \mathrm{d} R\,J_m(\alpha_{j,m}\,R/a)\allowbreak\int_0^{2\pi}\der\phi\,\ee^{-\ii m\phi}w_n(R,\phi,\omega_n)$ where $\alpha_{j,m}$ denotes $j$-th zero of the Bessel function of order $m$. After substituting Eq.~\eqref{Eq:un} into Eq.~\eqref{Eq:Gamma_lfNR}, we find
\begin{align}
    \Gamma_{\li}^{\lf}(\omega)&=\frac{2e^2 }{\hbar\, v^2}\sum_{n=1}^{\infty} g_n(\omega)\int_0^{\Qc} \Qf \, \mathrm{d} \Qf
    \left\lvert\int_0^{\Rcut} R \, \mathrm{d} R\,f_{\li}(R)\right.\nonumber\\
    &\left.\times J_{\lf}(\Qf R) \sum_{j=1}^{\infty} J_{\Delta\ell}\left(\alpha_{j,\Delta\ell}\frac{R}{a}\right)A_{j,-\Delta\ell}^n \right\rvert^2,
     \label{Eq:Gama_expansion}
\end{align}
which immediately shows that to observe the emergence of a dichroic signal, the coefficients $A_{j,\Delta\ell}$ have to be different for $+\Delta\ell$ and $-\Delta\ell$. For numerical calculations, we used $a=5R_0$, and cutoff radius $\Rcut=0.7R_0$. In the case of a nanohelix, we can proceed with the evaluation of the coefficients (see Sec.~B1 of SM \cite{SM}) if we approximate the distribution of charges corresponding to the resonant eigenmodes with a cosine dependence $\mathrm{cos}(n\phi/2)$. This approximate charge distribution yields an excellent agreement between analytically and numerically calculated $w_n$ when evaluated further away from the nanoparticle boundaries (see Fig.~S5 in SM \cite{SM}). The loss probability then becomes
\begin{widetext}
\begin{align}
    \Gamma^\mathrm{\lf}_{\li}(\omega)&=\frac{8e^2}{\hbar}\sum_{n=1}^{\infty} \frac{\mathcal{N}_n^2  g_n(\omega)}{v^2} \int_0^{\Qc} \Qf \, \mathrm{d} \Qf \left\lvert \sum_{j=1}^{\infty} \frac{J_{\Delta\ell}\left(\alpha_{j,\Delta\ell}\frac{R_0}{a}\right)+\alpha_{j,\Delta\ell}  I_{\Delta\ell}\left(\frac{\omega_n R_0}{v}\right) J_{\Delta\ell+1}(\alpha_{j,\Delta\ell})  K_{\Delta\ell}\left(\frac{\omega_n  a}{v}\right)}{\left(\alpha_{j,\Delta\ell}\right)^2+\left(\frac{\omega_n  a }{v}\right)^2}\right.\nonumber\\
    &\left.\times \int_0^{\Rcut} R \, \mathrm{d} R\,f_{\li}(R)  J_{\lf}(\Qf R) \frac{J_{\Delta\ell}\left(\alpha_{j,\Delta\ell}\frac{R}{a}\right)}{\left[J_{\Delta\ell+1}\left(\alpha_{j,\Delta\ell}\right)\right]^2} \right\rvert^2\underbrace{\left\lvert \frac{(1 + (-1)^{1 + n} \ee^{-\ii t_n}) (t_n/(2\pi)-\Delta\ell)}{(t_n /(2\pi)-\Delta\ell)^2 - (n/2)^2} \right\rvert ^2}_{\mathcal{F}_{\Delta\ell}\left(t_n\right)},
   \label{Eq:Gama_helix_analytical}
\end{align}
\end{widetext}
where $\mathcal{N}_n=\frac{\xi}{\sigma_{n}} \sqrt{R_0^2 + \left(\frac{d}{2\pi} \right)^2}$ with $\sigma_n$ being normalisation constant obtained from MNPBEM simulation (see Sec.~C2 of SM \cite{SM} for details), $2\xi$, $d$ and $R_0$ are thickness, pitch and radius of the nanohelix, respectively. $I_m$ and $K_m$ are modified Bessel functions of the first and second kind, respectively, and order $m$. More importantly, Eq.~\eqref{Eq:Gama_helix_analytical} [as well as Eq.~\eqref{Eq:Gamma_lfNR}] depends primarily on $\Delta\ell$ rather than the exact values of $\li$ and $\lf$. The sole dependence on the initial OAM is through the incident wave function's radial profile $f_{\li}$. It is also straightforward to see that there exist $t_n=\omega_n d/v$ for which the function $\mathcal{F}_{\Delta\ell}$ differs from $\mathcal{F}_{-\Delta\ell}$. This condition, indicating the emergence of dichroism, can be fulfilled for all modes.

\begin{figure}
    \centering
    \includegraphics[width=0.85\linewidth]{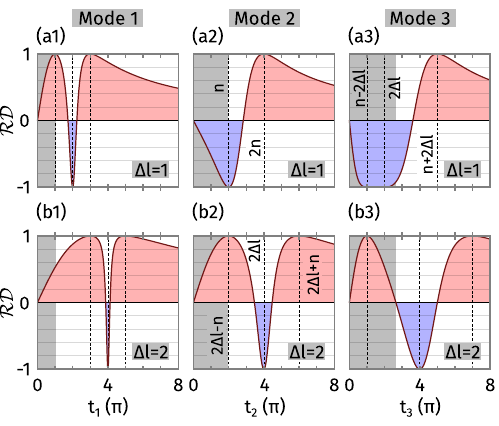}
    \caption{Relative dichroism dependence on the parameter $t_n(\omega_n,d,v)$ for the first three modes (each column corresponds to one mode). The upper row (a1--a3) is calculated for $\Delta\ell=1$ while the lower row (b1--b3) for $\Delta\ell=2$. Notice the similarities in panels (a1) and (b1--b3) and a qualitatively different behaviour in (a2) and (a3). The values of $t_n$ for maximal and minimal $\mathcal{RD}$ are displayed in (a2, a3), and (b2). Shaded regions depict relevant $t_n$ intervals for the first three modes studied in Fig.~\ref{fig:Fig2}(b1--d3).}
    \label{fig:RD(t)}
\end{figure}

To evaluate the dichroic signal in EELS, we first focus on the relative dichroism, defined as the relative difference between the EEL probability for a transition $\li\rightarrow\lf$ and its symmetric counterpart $-\li\rightarrow-\lf$, resulting in a compact expression when taking advantage of Eq.~\eqref{Eq:Gama_helix_analytical} and evaluating at energies of individual modes
\begin{align}
     \mathcal{RD}_{\li}^{\lf}(\omega_n) 
     &= \frac{{\mathcal{F}_{\Delta\ell}\left(t_n\right)}-{\mathcal{F}_{-\Delta\ell}\left(t_n\right)}}{{\mathcal{F}_{\Delta\ell}\left(t_n\right)}+{\mathcal{F}_{-\Delta\ell}\left(t_n\right)}}.
\label{Eq:Rel_dichroism_Fs}
\end{align}
The relative dichroism is particularly useful for comparing dichroic spectra among different initial parameters despite potentially large differences in spectral magnitudes. Furthermore, here it is solely governed by $\Delta \ell$ and $t_n$ within $\mathcal{F}_{\Delta\ell}(t_n)$.
By substituting for $\mathcal{F}_{\Delta\ell}(t_n)$ in Eq.~\eqref{Eq:Rel_dichroism_Fs} and considering only positive $\Delta\ell$ (negative $\Delta\ell$ yields opposite results) and small positive $t_n$ (corresponding to a small right-handed structures probed at tens of keV), we can analytically determine the values of $t_n$ that yield maximal $(\pm1)$ or zero relative dichroism. 

The resultant $t_n$-dependence is governed by the comparison of $n$ and $2\Delta\ell$.
For $n<2\Delta\ell$ [e.g., Fig.~\ref{fig:RD(t)}(a1,b1--b3)], the relative dichroism starts increasing to positive values, reaching its first maximum at $t_n=\pi(2\Delta\ell-n)$.
The change in sign of the relative dichroism shown in Fig.~\ref{fig:RD(t)} is a significant feature of the system. When studying helical particles with fixed geometry (constant $d$), maximal or minimal relative dichroism can be achieved by changing the acceleration voltage of the probe (i.e., changing the velocity $v$). It is therefore crucial to know in which region of the parametric space we probe. For $n>2\Delta\ell$ [e.g., Fig.~\ref{fig:RD(t)}(a3)] the behaviour is more complex. The relative dichroism starts decreasing to negative values, reaching its first minimum at $t_n=\pi(n-2\Delta\ell)$. Importantly, after this minimum, the relative dichroism does not return to zero but continues to a second minimum at $t_n=2\pi\Delta\ell$, forming a ``plateau''. For $n=3$ and $\Delta\ell=1$, this plateau results in the relative dichroism remaining very close to $-1$ over a wide range of $t_n$ parameters.
When $n = 2\Delta \ell$ [e.g., Fig.~\ref{fig:RD(t)}(a2)],
a special case with only two extrema (one minimum and one maximum) can occur.

\begin{figure}
    \centering
    \includegraphics[width=0.85\linewidth]{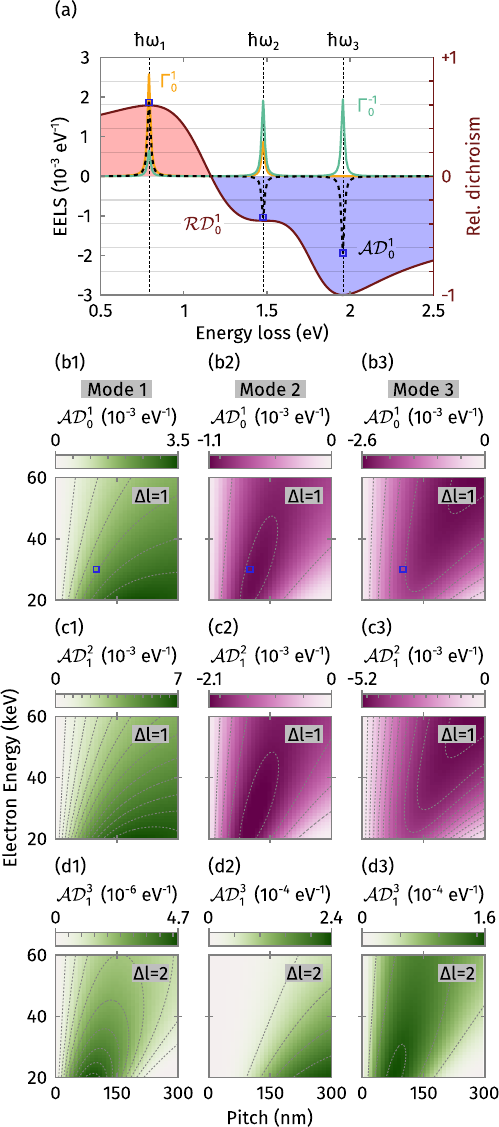}
    \caption{Dichroism for an infinitesimally thin silver single-twist helix [$R_0=50$~nm, a thickness of $\xi=10$~nm was used to determine the normalisation constant $\sigma_n$ from MNPBEM simulation, pitch is $d=100$~nm in (a) and varies in (b1--d3)] perfectly aligned with an electron beam [waist $\BW=$10~nm, energy is 30~keV in (a) 20--60~keV in (b--d)]. We consider the collection of all electrons, thus $\Qc\rightarrow\infty$. (a) EEL spectra for $\li=0$ and positive ($\Gamma_0^1$; yellow) and negative ($\Gamma_0^{-1}$; green) OAM exchange with corresponding absolute dichroism ($\mathcal{AD}_0^1$; dashed black) and relative dichroism ($\mathcal{RD}_0^1$; maroon). 
    (b) Absolute dichroism for the first three modes as a function of the helix pitch and electron energy for the OAM transition $0\rightarrow\pm1$. The contour step is $5\cdot10^{-4}\,\mathrm{eV}^{-1}$, and the blue rectangles indicate the values of the absolute dichroism in (a).
    (c) Same as (b) for $\li=1$ and the transition in OAM $1\rightarrow\pm2$. 
    (d) Same as (c) but for the OAM transition $1\rightarrow\pm3$. The contour steps are $5\cdot10^{-7}$, $5\cdot10^{-5}$ and $5\cdot10^{-5}\,\mathrm{eV}^{-1}$ respectively. Ticks in colorbars mark the contour levels.}
    \label{fig:Fig2}
\end{figure}

Although the relative dichroism is often used as a measure of the dichroic signal even in light-optics experiments, it has a significant drawback: it can yield maxima even for a very small spectral response. Particularly in EELS, which can yield poor signal-to-noise ratios (SNRs), it might be desirable to rely on the absolute dichroism defined as $\mathcal{AD}_{\li}^{\lf}(\omega) = \Gamma_{\li}^{\lf}(\omega) - \Gamma_{-\li}^{-\lf}(\omega)$. By substituting Eq.~\eqref{Eq:Gama_helix_analytical} and evaluating the $\mathcal{AD}$ at the energies of individual modes, we obtain
\begin{align}
    \mathcal{AD}_{\li}^{\lf}(\omega_n) &= \frac{8\,e^2\mathcal{N}_{n}^2 \left[{\mathcal{F}_{\Delta\ell}\left(t_n\right)}-{\mathcal{F}_{-\Delta\ell}\left(t_n\right)}\right]}{\hbar v^2}g_n(\omega_n) \nonumber\\
    &\times\int_0^{\Qc} \Qf \, \mathrm{d} \Qf \left\lvert \sum_{j=1}^{\infty} \mathcal{Z}_{j,\Delta\ell}\int_0^{\Rcut} R \, \mathrm{d} R\,f_{\li}(R) \right. \nonumber\\
    &\left.\times J_{\lf}(\Qf R)J_{\Delta\ell}\left(\alpha_{j,\Delta\ell}\frac{R}{a}\right)\right\rvert^2,
    \label{Eq:Abs_dich_final}
\end{align}
which is again strongly governed by the functions $\mathcal{F}_{\Delta\ell}\left(t_n\right)$, but also scaled by the integral featuring spatial dependence of the wave functions. For brevity, we used a function $\mathcal{Z}_{j,\Delta\ell}$ defined by Eq.~(S17) in SM \cite{SM}. It is clear from the definitions of dichroism in Eqs.~\eqref{Eq:Rel_dichroism_Fs}~and~\eqref{Eq:Abs_dich_final} that it can appear only if the electron transfers OAM to the sample, i.e. $\Delta\ell \neq 0$. 

We compare the relative and absolute dichroism for a silver thin helix in Fig.~\ref{fig:Fig2}(a). We assume the simplest experimental configuration, featuring an initially Gaussian electron beam with $\ell_i=0$. Selecting the lowest possible OAMs ($\lf=\pm1$) we obtain the EEL spectra for the positive ($\Gamma_{0}^{1}$) and negative ($\Gamma_{0}^{-1}$) difference of OAM, and corresponding absolute $\mathcal{AD}_{0}^{1}$ and relative $\mathcal{RD}_{0}^{1}$ dichroism. From Eq.~\eqref{Eq:Rel_dichroism_Fs} and for $\Delta\ell=1$ we see that the intensity of the relative dichroism for the first mode is always positive. On the other hand, it is negative for the second and third modes, which can be explained by the probed $t_n$ regions visualized as the shaded areas in Fig.~\ref{fig:RD(t)}(a1--a3).
The third mode exhibits $\mathcal{RD}$ very close to the value -1 for the investigated region of helix pitches and electron energies (see Fig.~S2(c3) for lower energies in SM \cite{SM}). By evaluating the parameter $t_3$, we can indeed find that we probe a large part of a ``plateau'' area in Fig.~\ref{fig:RD(t)}(a3), where the analytical model yields $\mathcal{RD}\approx -1$.

Next, we focus on the dependence of $\mathcal{AD}$ on some of the relevant experimental parameters, which can help determine whether the signal is measurable with one's experimental equipment. Fig.~\ref{fig:Fig2}(b1--b3) shows the intensity maps of $\mathcal{AD}_0^1$ as a function of the acceleration voltage and helix pitch for the first three modes [we note that the blue rectangles represent the peak values from Fig.~\ref{fig:Fig2}(a)]. Qualitatively following the predictions obtained from the analysis of $\mathcal{RD}$, the optimal parameters are different for each mode.

Similar results can be seen in Fig.~\ref{fig:Fig2}\mbox{(c1--c3)}, although the incident beam is now Laguerre-Gaussian with \mbox{$\li=1$}. Because the OAM difference is the same as in Fig.~\ref{fig:Fig2}\mbox{(b1--b3)}, the absolute dichroism is only scaled by a different value of the overlap integral in Eq.~\ref{Eq:Abs_dich_final}. In this case, $\mathcal{AD}_1^2$ is approximately twice as large compared to $\mathcal{AD}_0^1$. This behavior can be intuitively explained by the donut-like region with a large beam intensity emerging for the $\li=1$ beam, which is closer to the helix boundaries and thus excites the nanoparticle more efficiently. Further optimization of the resulting $\mathcal{AD}$ could be performed by varying beam convergence ($\BW$) and collection conditions. We also note that even though the $\mathcal{RD}\approx -1$ for $n=3$ and almost all pitches and electron energies under consideration [see Fig.~S2(c3)], the $\mathcal{AD}$ for low energies \mbox{(10--20~keV)} and large pitches (300~nm) is close to zero -- the dichroic signal would be very difficult to obtain experimentally [see Fig.~S2(a3)].

In Fig.~\ref{fig:Fig2}(d1--d3) we explore the case of $\Delta\ell=2$ by evaluating $\mathcal{AD}_1^3$, which reaches positive values for all three modes (we note that $\mathcal{AD}_0^2$ will behave similarly except for a different scaling). This can be easily predicted by looking at the shaded region in Fig.~\ref{fig:RD(t)}(b1--b3). Although the transition $1\rightarrow 3$ provides approximately an order of magnitude lower values of $\mathcal{AD}$ compared to the cases with $\Delta \ell=1$, it could be promising for the determination of the handedness of the chiral object. If the object is right-handed (resp. left-handed), the absolute (and relative) dichroism is positive (resp. negative) for the lowest three modes, and electron energies higher than 20~keV. 

\begin{figure}
    \centering
    \includegraphics[width=.8\linewidth]{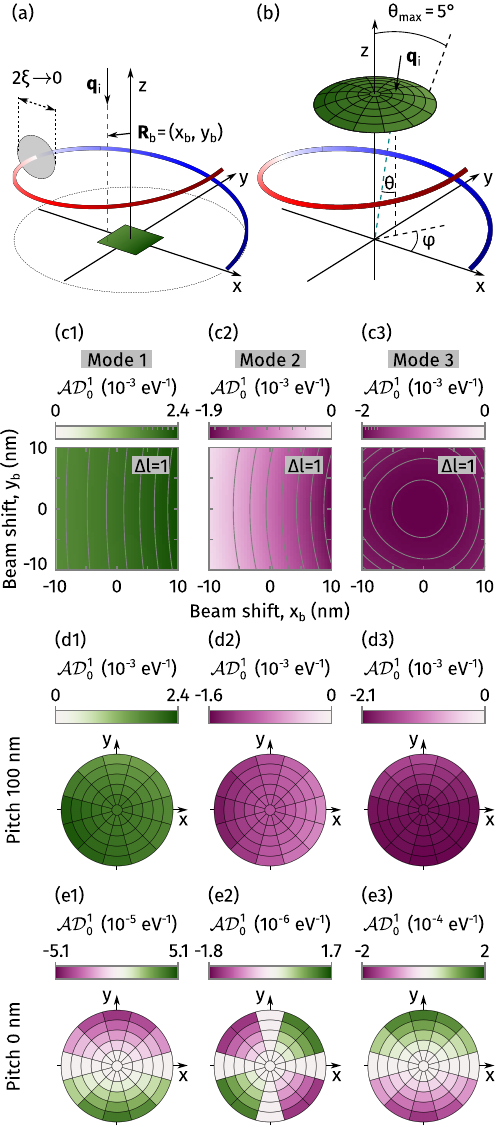}
    \caption{Absolute dichroism $\mathcal{AD}_{0}^{1}$ for an infinitesimally thin silver single-twist helix [$R_0= 50$~nm, $\xi=10$~nm, $d=100$~nm in (c,d) and $d=0$~nm in (e)] probed by a 30~keV Gaussian electron beam with $\BW=10$~nm, and $\alphaf=5$~mrad when the electron beam is scanned or tilted.
    (a) Scheme of the ``scanning'' simulation, where the beam's trajectory is parallel to the helix axis, impinging at position $\Rb_\mathrm{b}=(x_\mathrm{b}, y_\mathrm{b})$. 
     $\mathcal{AD}$ is plotted in the $20\times20$~nm computation area in (c1--c3) for the first three modes.
    The contour steps are $\{1;2;0.5\}\cdot10^{-4}\,\mathrm{eV}^{-1}$ respectively.
   (b) Scheme of the ``tilting'' simulation for the beam tilt represented in the spherical coordinate system by angles $(\theta;\varphi)$. The electron beam passes through the origin of the coordinate system.
   Each ``pixel'' in the circular ``lid'' represents a unique impinging direction of the beam. 
    The results are computed for polar angles $\theta$ ranging from $0\degree$ to $5\degree$ with a step $1\degree$. The azimuthal dependence is divided into 12 parts. 
    (d1--d3) $\mathcal{AD}_0^1$ for the first three modes when tilting the beam. 
    (e1--e3) Same as (c1--c3) but for a ``C-shaped'' structure ($d=0$~nm) exhibiting extrinsic dichroism.
    }
    \label{fig:Fig34}
\end{figure}

So far, we have considered an idealized scenario in which an electron beam passes exactly through the axis of the helix. However, one can suspect that under realistic experimental conditions, the beam can be misaligned -- shifted or tilted as sketched in Fig.~\ref{fig:Fig34}(a,b), respectively. We again utilize the analytical expression for the helix's eigenpotentials, but due to the disturbed symmetry, we evaluate Eq.~\eqref{Eq:Gamma_lfNR} numerically. Importantly, when probing the first three modes, the absolute dichroism $\mathcal{AD}_0^1$ only shows quantitative changes for both beam shift [Fig.~\ref{fig:Fig34}(c1--c3)] and tilt [Fig.~\ref{fig:Fig34}(d1--d3)]. These findings hold not just for the lowest transition with $\Delta\ell=1$ and an incident Gaussian beam ($\li=0$) (with $\BW=10$\,nm, $\alphaf=5$~mrad and energy 30\,keV), but also for other initial and final electron states (see Figs.~S3 and S1) and $\mathcal{RD}$ again proving that $\Delta\ell$ is the critical parameter determining behavior of the dichroic signal. However, in certain cases [see e.g., Fig.~\ref{fig:Fig34}(c2)], the dichroic signal can change quantitatively, which could be crucial when fighting with SNR in the experiment, and, more importantly, a nonzero dichroic signal can emerge even when probing an achiral nanostructure. If the system of a beam and a probed object, e.g., a ``C-shaped'' nanostructure (pitch $d=0$~nm) under a tilt is chiral, an ``extrinsic'' dichroism in EELS emerges as demonstrated in Fig.~\ref{fig:Fig34}(d1--d3).

We present a theoretical framework that allows for quantitative evaluation of the optical dichroic signal in electron energy-loss spectroscopy. Using an example of a plasmonic nanohelix, we demonstrate that both the sign and magnitude of the dichroic signal are governed not only by geometrical parameters of the sample and properties of its optical modes, but also by the exchanged OAM ($\Delta \ell$) and electron energy. We find analytical conditions to maximize the relative dichroism, analogous to phase-matching conditions for maximizing EELS with conventional electron beams \cite{DiGiulio2024,Zhao2025,Xie2025}, and demonstrate the importance of theoretical analysis by showing the complex dependence of the dichroic signal over the parameter space. 

We also focus on the influence of geometrical misalignments that may affect experiments. We show that, within a reasonable level of precision, the dichroic signal should be robust with respect to shifts and tilts between the sample and beam axes. However, it is still important to check the dependence of the signal on the tilt, as a non-zero, yet small and volatile, extrinsic dichroic signal can emerge for a tilted achiral sample. We further suggest that it is preferable to focus on low OAM transfers and small $\li$, ensuring the best spatial overlap with the projected sample potential, which typically varies on the nanoscale level when dealing with optical excitations. In such a scenario, one can achieve higher magnitudes of absolute dichroism, effectively overcoming issues with low signal-to-noise ratios. We note that the presented magnitudes could be further enhanced by optimizing the incident electron beam profile for an optimal overlap with the projected sample potential.

Our work sets the ground for near-future measurements of optical dichroism in setups that allow post-selection of electron states based on OAM. Such experiments could help to understand the OAM transfer and behavior of chiral optical fields at the nanoscale.

\begin{acknowledgments}
We acknowledge the support of the Czech Science Foundation GACR under the Junior Star grant No. 23-05119M. M.Z. acknowledges Brno Ph.D. Talent Scholarship - Funded by the Brno City Municipality.
\end{acknowledgments}

\bibliographystyle{apsrev4-1}

\bibliography{Bibliography_Main_SM.bib}



\clearpage
\begin{widetext}

\renewcommand{\theequation}{S\arabic{equation}}
\setcounter{equation}{0}
\renewcommand{\thefigure}{S\arabic{figure}}
\setcounter{figure}{0}
\noindent
\begin{center}\textbf{\MakeUppercase{Supplemental Material}}
\end{center}
\subsection{Theoretical framework for electron energy-loss spectroscopy with shaped (vortex) electron beams at optical frequencies}
\subsubsection{Description of electron wave functions}
The fast electrons moving along the $z$ axis can be described by wave functions taking the form
\begin{align}
    \psi_\mathrm{\lbrace i/f\rbrace }(R,\phi,z)=\frac{1}{\sqrt{L}}\mathrm{e}^{\mathrm{i}q_{z,\mathrm{\lbrace i/f\rbrace }} z}\psi_{\perp,\mathrm{\lbrace i/f\rbrace }}(R,\phi),
    \label{Eq:psii,R(f)}
\end{align}
where $(R,\phi,z)$ denote cylindrical coordinates, subscripts i/f stand for initial/final states, $\hbar q_{z,\mathrm{i}}$ and $\hbar q_{z,\mathrm{f}}$ are components of the initial/final electron momenta parallel with the beam axis, $L$ is the quantization length along $z$, and $\psi_{\perp,\mathrm{\lbrace i/f\rbrace }}$ describes dependence of the states in the transverse plane. We will consider a well-defined incident state in the form of a focused vortex beam i.e. Laguerre-Gauss beam:
\begin{align}
\psi_{\perp,\mathrm{i}}(R,\phi)=\ee^{\ii\li\phi} f_{\li}(R) = \ee^{\ii\li\phi}
\sqrt{\frac{p!}{\pi (p+\lvert \li\rvert)!}}\left(\frac{\sqrt{2}}{\BW}\right)^{\lvert \li\rvert+1} R^{\lvert \li\rvert} \ee^{-\frac{R^2}{\BW^2}}L_p^{\lvert \li\rvert}\left(\frac{2R^2}{\BW^2}\right),
     \label{Eq:psii}
\end{align}
where $\BW$ is the beam waist, $\hbar\li$ is orbital angular momentum (OAM) of the initial state with its quantum number~$l_\ii$ usually referred to as the (initial) topological charge. $L_p^{\lvert \ell\rvert}$~are associated Laguerre polynomials. For simplicity, we will consider only $p=0$ leading to $L_{0}^{\lvert\li\lvert}\left(x\right)=1$.

We consider a particular basis for the final states, which could be distinguished using an OAM analyzer, expressed as
\begin{align}
\psi_{\perp,\mathrm{f}}(R,\phi)=\frac{1}{\sqrt{A}}\mathrm{e}^{\mathrm{i}\lf\phi}J_{\lf}(Q_\mathrm{f}R),
\label{Eq:psi_fperp}
\end{align}
where $J_\ell$ are Bessel functions of order $\ell$, $\Qf$ is radial component of the final-state wavevectors, $\hbar\lf$ describes final OAM, and $A$ is the quantization area in the $x$-$y$ transversal plane.

\subsubsection{Non-retarded loss probability}
The loss probability corresponding to a transition from an initial electron state to states as described above can be expressed as \cite{GDA}
\begin{align}
\Gamma^\mathrm{NR}(\omega)=\frac{2e^2L}{\hbar v}\sum_\mathrm{f}\int\mathrm{d}^3\rb\int\mathrm{d}^3\rb'\,\psi^\ast_\mathrm{i}(\rb)\psi_\mathrm{i}(\rb')\psi^\ast_\mathrm{f}(\rb')\psi_\mathrm{f}(\rb)\,\mathrm{Im}\left\lbrace -W(\rb,\rb',\omega)\right\rbrace \delta(\epsilon_\mathrm{f}-\epsilon_\mathrm{i}+\omega),
\label{Eq:GammaNR1}
\end{align}
where $e$ is the elementary charge, $\me$ is the electron mass, $\hbar\epsilon_\mathrm{f/i}$ is the final/initial electron energy, $v$ is the electron velocity, and $W$ is the screened potential introducing the response of the probed structure in the non-retarded approximation. The particular form of electron states allows separability in terms of longitudinal and transverse momenta. The collection over final states can thus be rewritten as $\sum_\mathrm{f}\rightarrow L/(2\pi)\int\mathrm{d}q_{z,\mathrm{f}}\sum_{\perp,\mathrm{f}}$. Furthermore, the non-recoil approximation allows to perform the integration over the final wavevectors yielding
\begin{align}
\int \mathrm{d} q_{z,\mathrm{f}}\, \ee^{\ii(q_{z, \mathrm{f}}-q_{z,\ii})(z-z')}\delta(\epsilon_\mathrm{f}-\epsilon_\ii+\omega)=\frac{1}{v} \ee^{-\ii\omega(z-z')/v}
\end{align}
and in turn to rewrite Eq.~\eqref{Eq:GammaNR1} as
\begin{align}
\Gamma^\mathrm{NR}(\omega)=\frac{e^2}{\pi\hbar v^2}\sum_\mathrm{\perp,f}\int\mathrm{d}^2\Rb\int\mathrm{d}^2\Rb'\,\psi^\ast_\mathrm{\perp,i}(\Rb)\psi_\mathrm{\perp,i}(\Rb')\psi^\ast_\mathrm{\perp,f}(\Rb')\psi_\mathrm{\perp,f}(\Rb)\, \mathcal{W}(\Rb,\Rb',\omega),
\label{Eq:GammaNR2}
\end{align}
where we defined the projected screened potential, which, importantly, becomes separable

\begin{align}
    \mathcal{W}(\Rb,\Rb',\omega)&=\int\mathrm{d}z\int\mathrm{d}z'\, { \mathrm{Im}\left\lbrace - W(\rb,\rb',\omega)\right\rbrace\ee^{-\ii\omega(z-z')/v}}\nonumber\\
    &=\sum_{n=1}^{\infty} g_n(\omega)w_n(\Rb,\omega)w_n^\ast(\Rb',\omega)\approx \sum_{n=1}^{\infty} g_n(\omega)w_n(\Rb,\omega_n)w_n^\ast(\Rb',\omega_{n}),
   \label{Eq:W,eigenmodes}
\end{align}
where $n$ labels eigenmodes of the quasi-static optical response of the probed nanostructure, $g_n$ dominantly determines the spectral response of each eigenmode, and the functions $w_n$ introduce spatial dependence of the response. In the following, we neglect the frequency dependence of the functions $w_n$ and evaluate them around the resonant frequency corresponding to each mode, $\omega_{n}$, which will be valid for narrow resonances and for spectrally non-overlapping modes. \\

Substituting Eqs.~\eqref{Eq:W,eigenmodes} and \eqref{Eq:psi_fperp} in Eq.~\eqref{Eq:GammaNR2}, considering a detector imposing a cutoff of final transverse wavevectors $\Qc$ due to a finite collection angle, which allows us to write $\sum_\mathrm{\perp, f} \rightarrow {A}/(2 \pi) \int_0^{\Qc} \Qf \, \mathrm{d} \Qf$, the loss probability corresponding to a transition from an initial state to a final state with a given $\lf$ becomes 
\begin{align}
    \Gamma^\mathrm{\lf}_{\li}(\omega)&=\frac{e^2 }{2\pi^2\hbar v^2}\sum_{n=1}^{\infty} g_n(\omega)\int_0^{\Qc} \Qf \, \mathrm{d} \Qf \int R \, \mathrm{d} R\,\int_0^{2\pi}\mathrm{d}\phi\,\ee^{\ii(\lf-\li)\phi}f_{\li}^\ast(R)J_{\lf}(\Qf R)w_n(R, \phi,\omega_n)\nonumber\\
    &\times \int R'\,\mathrm{d}R'\,\int_0^{2\pi}\mathrm{d}\phi'\, \ee^{-\ii(\lf-\li)\phi'}f_{\li}(R')J_{\lf}(\Qf R')  w_n^\ast(R', \phi',\omega_n)\nonumber\\
    &=\frac{e^2 }{2\pi^2\hbar v^2}\sum_{n=1}^{\infty} g_n(\omega)\int_0^{\Qc} \Qf \, \mathrm{d} \Qf \left\vert\int R \, \mathrm{d} R\,\int_0^{2\pi}\mathrm{d}\phi\,\ee^{\ii\Delta\ell\phi}f_{\li}(R)J_{\lf}(\Qf R)w_n(R, \phi,\omega_n) \right\vert^2
\end{align}
which corresponds to Eq.~(1) of the main text and where we denoted $\Delta \ell=\lf-\li$ and considered the detector centered at $R=0$. Also, in all our cases, the function $f_{\li}(R)$ describing the radial dependence is real-valued throughout the whole space; thus, we set $f_{\li}=f_{\li}^\ast$.

We further express the spatial characteristics of the modes on the basis of suitable symmetry. If we consider a finite radius $a$, where we evaluate the projected potential, which is related to the natural spatial extent of near fields corresponding to the localized modes, we have Eq.~(2) of the main text:
\begin{align}
w_n(R,\phi,\omega_{n})=\sum_{m=-\infty}^\infty\ii^m\ee^{\ii m\phi}\sum_{j=1}^\infty J_m\left(\alpha_{j,m}\frac{R}{a}\right)A_{j,m}^n,
\end{align}
where $m$ are integers, and $A_{j,m}^n$ are (complex) expansion coefficients evaluated as
\begin{align*}
    A_{j,m}^n=\frac{(-\ii)^m}{\pi a^2 \left[J_{m+1}\left(\alpha_{j,m}\right)\right]^2} \int_0^a R \, \mathrm{d} R\,J_m\left(\alpha_{j,m}\frac{R}{a}\right)\int_0^{2\pi}\der\phi\,\ee^{-\ii m\phi}w_n(R,\phi,\omega_n),
\end{align*}
where $\alpha_{j,m}$ is $j$-th zero of the Bessel function $J_m$ of order $m$. After substitution of Eq.~(2) into Eq.~(1), we obtain

\begin{align}
    \Gamma^\mathrm{\lf}_{\li}(\omega)&=\frac{e^2 }{2\pi^2\hbar v^2}\sum_{n=1}^{\infty} g_n(\omega)\int_0^{\Qc} \Qf \, \mathrm{d} \Qf \left\lvert\int_0^{\Rcut} R \, \mathrm{d} R\,f_{\li}(R)J_{\lf}(\Qf R) \sum_{m=-\infty}^\infty\sum_{j=1}^{\infty} \ii^mJ_m\left(\alpha_{j,m}\frac{R}{a}\right)A_{j,m}^n \int_0^{2\pi}\mathrm{d}\phi\,\ee^{\ii(\Delta \ell+m)\phi}\right\rvert^2\nonumber\\
     &=\frac{2e^2 }{\hbar v^2}\sum_{n=1}^{\infty} g_n(\omega)\int_0^{\Qc} \Qf \, \mathrm{d} \Qf \left\lvert\int_0^{\Rcut} R \, \mathrm{d} R\,f_{\li}(R)J_{\lf}(\Qf R) \sum_{j=1}^{\infty} J_{-\Delta\ell}\left(\alpha_{j,-\Delta\ell}\frac{R}{a}\right)A_{j,-\Delta\ell}^n \right\rvert^2.
     \nonumber \\
     &=\frac{2e^2 }{\hbar v^2}\sum_{n=1}^{\infty} g_n(\omega)\int_0^{\Qc} \Qf \, \mathrm{d} \Qf \left\lvert\int_0^{\Rcut} R \, \mathrm{d} R\,f_{\li}(R)J_{\lf}(\Qf R) \sum_{j=1}^{\infty} J_{\Delta\ell}\left(\alpha_{j,\Delta\ell}\frac{R}{a}\right)A_{j,-\Delta\ell}^n \right\rvert^2.
     \label{Eq:Gama_expansion2}
\end{align}
To observe the emergence of dichroic signal, the coefficients $A_{j,\Delta\ell}$ have to be different for $\pm\Delta\ell$. 

\subsection{Non-retarded loss probability for a single twist helix}
\subsubsection{Analytical evaluation of the functions $w_n$}
\label{Sec:Analytical_approx_coef}
For some simple geometries, the screened potential can be evaluated analytically, but generally, a numerical solution must be found. We first study a simplified case of a single-twist helix approximated by a line charge distribution. If we consider an infinitely thin wire forming the single-twist helix centered at the origin, we can associate the eigenmodes formed along the helix with the volume charge density
\begin{equation}
    \rho_n(R,\phi,z)=\frac{2 \pi \xi}{\sigma_{n}} \sqrt{R_0^2 + \left(\frac{d}{2\pi} \right)^2}\mathrm{cos}\left(\frac{n\phi}{2}\right)\frac{\delta(R-R_0)}{R}\delta[z-\phi d/(2\pi)],
\end{equation}
where $R_0$ is the radius, $d$ is the pitch and $n$ is the mode index. $\sigma_{n}$ is the MNPBEM norm (see Sec.~\ref{Sec:EigenmodeNormalization}) and $\xi$ is the radius of the wire forming the helix (necessary for the transition from the 2D charge density employed in MNPBEM to the 1D charge density used here). The function $w_n$---calculated according to Eq.~\eqref{Eq:wnProjectedSI}---is then
\begin{align}
    w_n(R,\phi,\omega ) & = \frac{\xi}{\sigma_{n}} \sqrt{R_0^2 + \left(\frac{d}{2\pi} \right)^2} \int_{0}^{2\pi} \mathrm{d}\phi^\prime \cos\left(\frac{n \phi^{\prime}}{2} \right) \, e^{- i \omega_n \phi^{\prime} d /(2\pi v)} K_{0} \left( \frac{\omega_n}{v} \sqrt{R^2+R_0^2 -2 R R_{0} \cos(\phi - \phi^{\prime}) } \right),
\end{align}
which yields the expansion coefficients

\begin{align}
    A_{j,m}^n&=\frac{\mathcal{N}_n(-\ii)^m}{\pi a^2 \left[J_{m+1}\left(\alpha_{j,m}\right)\right]^2}\int_0^a R \, \mathrm{d} R\,J_m\left(\alpha_{j,m}\frac{R}{a}\right)\nonumber\\
    &\times\int_0^{2\pi}\der\phi\,\ee^{-\ii m\phi}\int_0^{2\pi}\der\phi'\,\mathrm{cos}(n\phi'/2)\ee^{-\ii\omega_n\phi'd/(2\pi v)}K_0\left(\frac{\omega_n\sqrt{R^2+R_0^2-2RR_0\mathrm{cos}(\phi-\phi')}}{v}\right)\nonumber\\
    &=\frac{2\mathcal{N}_n(-\ii)^m}{ a^2 \left[J_{m+1}\left(\alpha_{j,m}\right)\right]^2}\int_0^a R \, \mathrm{d} R\,J_m\left(\alpha_{j,m}\frac{R}{a}\right)I_m\left(\frac{\omega_n R_<}{v}\right)K_m\left(\frac{\omega_n R_>}{v}\right)\underbrace{\int_0^{2\pi}\der\phi'\,\mathrm{cos}(n\phi'/2)\ee^{-\ii\omega_n\phi'd/(2\pi v)}\ee^{-\ii m\phi'}}_{A_\phi}\nonumber\\
    &=\frac{2\mathcal{N}_n(-\ii)^m}{\left[J_{m+1}\left(\alpha_{j,m}\right)\right]^2}\frac{J_{m}\left(\alpha_{j,m}\frac{R_0}{a}\right)+\alpha_{j,m}I_{m}\left(\frac{\omega_n R_0}{v}\right)J_{m+1}(\alpha_{j,m})K_{m}\left(\frac{\omega_n a}{v}\right)}{\left(\alpha_{j,m}\right)^2+\left(\frac{a\omega_n}{v}\right)^2}\nonumber\\
    &\times
     \left\{ \begin{array}{c l}
    \pi&\mathrm{when}\quad n/2=\pm(\omega_n d/(2\pi v)+m)\\
    \frac{-\ii(1 + (-1)^{1 + n} \ee^{-\ii \omega_n d/ v}) (\omega_n d/(2\pi v)+m)}{(\omega_n d/(2\pi v)+m)^2 - (n/2)^2}&\mathrm{otherwise},
     \end{array}\right.
    \label{Eq:helix_analytical}
\end{align}
where $R_<=\mathrm{min}\lbrace R,R_0\rbrace$, $R_>=\mathrm{max}\lbrace R,R_0\rbrace$ and $\mathcal{N}_n=\frac{\xi}{\sigma_{n}} \sqrt{R_0^2 + \left(\frac{d}{2\pi} \right)^2}$.

By evaluating the azimuthal integral $(A_{\phi})$ above we obtain the basis of the term $\mathcal{F}_{\Delta\ell}(t_n)$. This integral depends only on the azimuthal distribution of the linear charge density $\cos(n\phi/2)$, the exponential term describing the change of phase during the electron passage, and another exponential term from the Fourier series. However, for the helical geometry, one can evaluate the generalized expression $\mathcal{F}'$ for an arbitrary azimuthal charge distribution $\tau(\phi)$ if we express the function $\tau(\phi)$ as a power series:
\begin{align*}
    A_{\phi}&=\int_0^{2\pi}\der\phi\,\tau(\phi)\ee^{-\ii\omega_n\phi'd/(2\pi v)}\ee^{-\ii m\phi} = \sum_{i=0}^{\infty} a_i \int_0^{2\pi}\der\phi\,\phi^i\ee^{-\ii\Xi\phi} = \sum_{i=0}^{\infty} a_i \left(\frac{1}{\ii\Xi}\right)^{i+1}\int_0^{2\pi\ii\Xi}\der t\,t^i\ee^{-t}\\
    &=\sum_{i=0}^{\infty} a_i \left(\frac{1}{\ii\Xi}\right)^{i+1} \gamma(i+1, 2\pi\ii\Xi) = \sum_{i=0}^{\infty} a_i \left(\frac{-2\pi\ii}{t_n+2\pi m}\right)^{i+1} \gamma\left(i+1, \ii t_n+2\pi\ii m\right)
\end{align*}
where $\Xi = \frac{\omega_nd}{2\pi v}+m$ and $\gamma(s,z)$ is the lower incomplete gamma function \cite[\href{http://dlmf.nist.gov/8.2.1}{(8.2.1)}]{NIST:DLMF}.
When we plug this expression into the Eq.~\eqref{Eq:helix_analytical} and then to Eq.~\eqref{Eq:Gama_expansion2}, the coefficient $m$ is changed to $-\Delta\ell$. $A_{\phi}$ is different for $\pm\Delta\ell$ for every $i$ and the $\left\lvert A_{\phi}\right\rvert^2$ expression is a generalized form of $\mathcal{F}_{\Delta\ell}(t_n)$.

\subsubsection{Loss probability for a single-twist helix probed by vortex electron beams}
After substitution of Eq.~\eqref{Eq:helix_analytical} into Eq.~(3), we can readily obtain Eq.~(4) in the main text:
{
\begin{align}
    \Gamma^\mathrm{\lf}_{\li}(\omega)&=\frac{8e^2}{\hbar}\sum_{n=1}^{\infty} \frac{\mathcal{N}_n^2g_n(\omega)}{v^2} \times\int_0^{\Qc} \Qf \, \mathrm{d} \Qf \left\lvert \sum_{j=1}^{\infty} \frac{J_{\Delta\ell}\left(\alpha_{j,\Delta\ell}\frac{R_0}{a}\right)+\alpha_{j,\Delta\ell}I_{\Delta\ell}\left(\frac{\omega_n R_0}{v}\right)J_{\Delta\ell+1}(\alpha_{j,\Delta\ell})K_{\Delta\ell}\left(\frac{\omega_n a}{v}\right)}{\left(\alpha_{j,\Delta\ell}\right)^2+\left(\frac{a\omega_n}{v}\right)^2}\right.\nonumber\\
    &\left.\times \int_0^{\Rcut} R \, \mathrm{d} R\,f_{\li}(R)J_{\lf}(\Qf R) \frac{J_{\Delta\ell}\left(\alpha_{j,\Delta\ell}\frac{R}{a}\right)}{\left[J_{\Delta\ell+1}\left(\alpha_{j,\Delta\ell}\right)\right]^2} \right\rvert^2\underbrace{\left\lvert \frac{(1 + (-1)^{1 + n} \ee^{-\ii \omega_n d/ v}) (\omega_n d/(2\pi v)-\Delta\ell)}{(\omega_n d/(2\pi v)-\Delta\ell)^2 - (n/2)^2} \right\rvert ^2}_{\mathcal{F}_{\Delta\ell}\left(\frac{\omega_n d}{v}\right)}.
   \label{Eq:Gama_helix_analytical2}
\end{align}
}
where we define the function $\mathcal{F}_{\Delta\ell}$ which is dependent only on the difference of topological charges $\Delta\ell$, mode index $n$, geometry of the sample -- helix pitch $d$, and the probe -- velocity of the electron $v$. From Eq.~\eqref{Eq:Gama_helix_analytical2} it is straightforward to see, that there exists $t_n=\omega_nd/v$ for which the function $\mathcal{F}_{\Delta\ell}(t_n)$ differs from $\mathcal{F}_{-\Delta\ell}(t_n)$. This condition holds for all modes, indicating dichroism.

To find the absolute and relative dichroism, we also need to calculate the loss probability for a flipped sign of $\Delta\ell$. We use Eq.~\eqref{Eq:Gama_helix_analytical2} and obtain:
\begin{align*}
    \Gamma^\mathrm{-\lf}_{-\li}(\omega)
    &=\frac{8e^2}{\hbar}\sum_{n=1}^{\infty} \frac{\mathcal{N}_n^2g_n(\omega)}{v^2}\int_0^{\Qc} \Qf \, \mathrm{d} \Qf \left\lvert \sum_{j=1}^{\infty} \frac{J_{-\Delta\ell}\left(\alpha_{j,-\Delta\ell}\frac{R_0}{a}\right)+\alpha_{j,-\Delta\ell}I_{-\Delta\ell}\left(\frac{\omega_n R_0}{v}\right)J_{-\Delta\ell+1}(\alpha_{j,-\Delta\ell})K_{-\Delta\ell}\left(\frac{\omega_n a}{v}\right)}{\left(\alpha_{j,-\Delta\ell}\right)^2+\left(\frac{a\omega_n}{v}\right)^2}\right.\nonumber\\
    &\left.\times\int_0^{\Rcut} R \, \mathrm{d} R\,f_{-\li}(R)J_{-\lf}(\Qf R) \frac{J_{-\Delta\ell}\left(\alpha_{j,-\Delta\ell}\frac{R}{a}\right)}{\left[J_{-\Delta\ell+1}\left(\alpha_{j,-\Delta\ell}\right)\right]^2} \right\rvert^2 \mathcal{F}_{-\Delta\ell}\left(t_n\right).
\end{align*}
For the zeros of Bessel functions $\alpha_{j,\ell}=\alpha_{j,-\ell}$ holds. Also, we can use relations \cite[\href{http://dlmf.nist.gov/10.4.E1}{(10.4.1)}]{NIST:DLMF}, \cite[\href{http://dlmf.nist.gov/10.6.E1}{(10.6.1)}]{NIST:DLMF} and \cite[\href{https://dlmf.nist.gov/10.27.E1}{(10.27.1)}]{NIST:DLMF} to find that $J_{-\ell+1}(\alpha_{j,\ell})=(-1)^{\ell}J_{\ell+1}(\alpha_{j,\ell})$ and obtain
\begin{align}
    \Gamma^\mathrm{-\lf}_{-\li}(\omega)
    &=\frac{8e^2}{\hbar}\sum_{n=1}^{\infty} \frac{\mathcal{N}_n^2g_n(\omega)}{v^2}\int_0^{\Qc} \Qf \, \mathrm{d} \Qf \left\lvert \sum_{j=1}^{\infty} \frac{(-1)^{\Delta\ell}J_{\Delta\ell}\left(\alpha_{j,\Delta\ell}\frac{R_0}{a}\right)+(-1)^{\Delta\ell}\alpha_{j,\Delta\ell}I_{\Delta\ell}\left(\frac{\omega_n R_0}{v}\right)J_{\Delta\ell+1}(\alpha_{j,\Delta\ell})K_{\Delta\ell}\left(\frac{\omega_n a}{v}\right)}{\left(\alpha_{j,\Delta\ell}\right)^2+\left(\frac{a\omega_n}{v}\right)^2}\right.\nonumber\\
    &\left.\times\int_0^{\Rcut} R \, \mathrm{d} R\,f_{-\li}(R)(-1)^{\lf}J_{\lf}(\Qf R) (-1)^{\Delta\ell}\frac{(-1)^{\Delta\ell}J_{\Delta\ell}\left(\alpha_{j,\Delta\ell}\frac{R}{a}\right)}{\left[(-1)^{\Delta\ell}J_{\Delta\ell+1}\left(\alpha_{j,\Delta\ell}\right)\right]^2} \right\rvert^2  \mathcal{F}_{-\Delta\ell}\left(t_n\right)\nonumber\\
    &=\frac{8e^2}{\hbar}\sum_{n=1}^{\infty} \frac{\mathcal{N}_n^2g_n(\omega)}{v^2} \times\int_0^{\Qc} \Qf \, \mathrm{d} \Qf \left\lvert \sum_{j=1}^{\infty} \frac{J_{\Delta\ell}\left(\alpha_{j,\Delta\ell}\frac{R_0}{a}\right)+\alpha_{j,\Delta\ell}I_{\Delta\ell}\left(\frac{\omega_n R_0}{v}\right)J_{\Delta\ell+1}(\alpha_{j,\Delta\ell})K_{\Delta\ell}\left(\frac{\omega_n a}{v}\right)}{\left(\alpha_{j,\Delta\ell}\right)^2+\left(\frac{a\omega_n}{v}\right)^2}\right.\nonumber\\
    &\left.\times \int_0^{\Rcut} R \, \mathrm{d} R\,f_{\li}(R)J_{\lf}(\Qf R) \frac{J_{\Delta\ell}\left(\alpha_{j,\Delta\ell}\frac{R}{a}\right)}{\left[J_{\Delta\ell+1}\left(\alpha_{j,\Delta\ell}\right)\right]^2} \right\rvert^2 \mathcal{F}_{-\Delta\ell}\left(t_n\right).
    \label{Eq:Gama_helix_analytical_negative}
\end{align}
Notice that the radial component of the incident states $f_{\li}(R)$ is invariant for flipped sign of $\li$ [see Eq.\eqref{Eq:psii}].

Eq.~\eqref{Eq:Gama_helix_analytical2} can be further simplified if we consider $\Qc\rightarrow\infty$:
\begin{align}
    \Gamma^\mathrm{\lf}_{\li}(\omega)&=\frac{8e^2}{\hbar}\sum_{n=1}^{\infty} \frac{\mathcal{N}_n^2g_n(\omega)}{v^2}\left\lvert\frac{(1 + (-1)^{1 + n} \ee^{-\ii \omega_n d/ v}) (\omega_n d/(2\pi v)-\Delta\ell)}{(\omega_n d/(2\pi v)-\Delta\ell)^2 - (n/2)^2}\right\rvert^2  \nonumber\\
    &\times\int_0^{\Rcut} R \mathrm{d} R\,f_{\li}^2(R) \left(\sum_{j=1}^{\infty} \frac{J_{\Delta\ell}\left(\alpha_{j,\Delta\ell}\frac{R_0}{a}\right)+\alpha_{j,\Delta\ell}I_{\Delta\ell}\left(\frac{\omega_n R_0}{v}\right)J_{\Delta\ell+1}(\alpha_{j,\Delta\ell})K_{\Delta\ell}\left(\frac{\omega_n a}{v}\right)}{\left(\alpha_{j,\Delta\ell}\right)^2+\left(\frac{a\omega_n}{v}\right)^2}  \frac{J_{\Delta\ell}\left(\alpha_{j,\Delta\ell}\frac{R}{a}\right)}{\left[J_{\Delta\ell+1}\left(\alpha_{j,\Delta\ell}\right)\right]^2} \right)^2\nonumber\\
    &=\frac{8e^2}{\hbar}\sum_{n=1}^{\infty} \frac{\mathcal{N}_n^2g_n(\omega)}{v^2}\mathcal{F}_{\Delta\ell}(t_n)\int_0^{\Rcut} R \mathrm{d} R\,f_{\li}^2(R)\left(\sum_{j=1}^{\infty} \mathcal{Z}_{j,\Delta\ell}J_{\Delta\ell}\left(\alpha_{j,\Delta\ell}\frac{R}{a}\right) \right)^2.
   \label{Eq:Gama_helix_analytical_simplified}
\end{align}
where we used \cite[\href{https://dlmf.nist.gov/10.22.E67}{(10.22.67)}]{NIST:DLMF} and defined
\begin{align}
    \label{Eq:function_Z}
   \mathcal{Z}_{j,\Delta\ell}=
    \frac{J_{\Delta\ell}\left(\alpha_{j,\Delta\ell}\frac{R_0}{a}\right)+\alpha_{j,\Delta\ell}I_{\Delta\ell}\left(\frac{\omega_n R_0}{v}\right)J_{\Delta\ell+1}(\alpha_{j,\Delta\ell})K_{\Delta\ell}\left(\frac{\omega_n a}{v}\right)}{\left(\alpha_{j,\Delta\ell}\right)^2+\left(\frac{a\omega_n}{v}\right)^2}  \frac{1}{\left[J_{\Delta\ell+1}\left(\alpha_{j,\Delta\ell}\right)\right]^2}.
\end{align}
Also notice that $\mathcal{Z}_{j,\Delta\ell} =\left(-1\right)^{-\Delta\ell} \mathcal{Z}_{j,-\Delta\ell}.$

Now, we can formulate the absolute and relative dichroism for the OAM transition $\hbar\Delta\ell$ when evaluated at a specific mode energy $\omega_n$ as
\begin{align}
    \mathcal{AD}_{\li}^{\lf}(\omega_n) &= \Gamma_{\li}^{\lf}(\omega_n) - \Gamma_{-\li}^{-\lf}(\omega_n)\nonumber\\
    &=\frac{8e^2\mathcal{N}_{n}^2 \left[{\mathcal{F}_{\Delta\ell}\left(t_n\right)}-{\mathcal{F}_{-\Delta\ell}\left(t_n\right)}\right]}{\hbar v^2}g_n(\omega_n)\int_0^{\Qc} \Qf \, \mathrm{d} \Qf \left\lvert \sum_{j=1}^{\infty} \mathcal{Z}_{j,\Delta\ell}\int_0^{\Rcut} R \, \mathrm{d} R\,f_{\li}(R) J_{\lf}(\Qf R)J_{\Delta\ell}\left(\alpha_{j,\Delta\ell}\frac{R}{a}\right)\right\rvert^2,
    \\
    \mathcal{RD}_{\li}^{\lf}(\omega_n)&= \frac{\Gamma_{\li}^{\lf}(\omega_n) - \Gamma_{-\li}^{-\lf}(\omega_n)}{\Gamma_{\li}^{\lf}(\omega_n) + \Gamma_{-\li}^{-\lf}(\omega_n)}=\frac{{\mathcal{F}_{\Delta\ell}\left(t_n\right)}-{\mathcal{F}_{-\Delta\ell}\left(t_n\right)}}{{\mathcal{F}_{\Delta\ell}\left(t_n\right)}+{\mathcal{F}_{-\Delta\ell}\left(t_n\right)}},
\end{align}
which corresponds to Eqs.~(6) and (5) of the main text.

\subsubsection{Loss probability for a single-twist helix and a plane-wave to vortex transition}

If we consider the simplest experiment consisting of illuminating the single-twist helix with an electron plane wave (featuring $\li=0$) and post-selecting some final OAM $\hbar\lf$, we obtain the loss probability
\begin{align}
     \Gamma_{\mathrm{PW}}^{\lf}(\omega)&=\frac{4e^2}{\hbar}\sum_{n=1}^{\infty} \frac{\mathcal{N}_n^2g_n(\omega)a^2}{v^2 A}\mathcal{F}_{\lf}(t_n)\sum_{j=1}^{\infty} \mathcal{Z}_{j, \lf}^2 \left[J_{\lf+1}(\alpha_{j,\lf}) \right]^2,
\end{align}
and absolute dichroism becomes
\begin{align}
    \mathcal{AD}_{\mathrm{PW}}^{\lf}(\omega)&=\frac{4e^2}{\hbar}\left[\mathcal{F}_{\lf}(t_n)-\mathcal{F_{-\lf}}(t_n)\right]\sum_{n=1}^{\infty} \frac{\mathcal{N}_n^2g_n(\omega)a^2}{v^2 A}\sum_{j=1}^{\infty} \mathcal{Z}_{j, \lf}^2 \left[J_{\lf+1}(\alpha_{j,\lf}) \right]^2,
\end{align}
where we integrated over the region where the screened potential is evaluated, thus $\Rcut=a$.

\subsubsection{Loss probability for a single-twist helix and a non-vortex focused beam to vortex transition}
By inserting the Laguerre-Gaussian radial profile in Eq.~\eqref{Eq:psii} with $\li=0$ and $p=0$ into Eq.~\eqref{Eq:Gama_helix_analytical_simplified} (for all electrons collected at the detector), we obtain the transition probability corresponding to a focused non-vortex beam to a vortex state
\begin{align}
    \Gamma^\mathrm{\lf}_{0}(\omega)&=\frac{8e^2}{\hbar}\sum_{n=1}^{\infty} \frac{\mathcal{N}_n^2g_n(\omega)}{v^2}  \mathcal{F}_{\lf}\left(t_n\right)\frac{2}{\pi\BW^2}
    \int_0^{\infty} R \mathrm{d} R\, \ee^{-\frac{2R^2}{\BW^2}}\sum_{j=1}^{\infty}\mathcal{Z}_{j,\lf}J_{\lf}\left(\alpha_{j,\lf}\frac{R}{a}\right) \sum_{k=1}^{\infty}\mathcal{Z}_{k,\lf}J_{\lf}\left(\alpha_{k,\lf}\frac{R}{a}\right)\nonumber\\
    &=\frac{8e^2}{\hbar}\sum_{n=1}^{\infty} \frac{\mathcal{N}_n^2g_n(\omega)}{v^2} \mathcal{F}_{\lf}\left(t_n\right)\frac{2}{\pi\BW^2}\sum_{j=1}^{\infty}\sum_{k=1}^{\infty}\mathcal{Z}_{j,\lf}\mathcal{Z}_{k,\lf}\int_0^{\infty} R \mathrm{d} R\,\ee^{-\frac{2R^2}{\BW^2}}J_{\lf}\left(\alpha_{j,\lf}\frac{R}{a}\right)J_{\lf}\left(\alpha_{k,\lf}\frac{R}{a}\right)\nonumber\\
    &=\frac{8e^2}{\hbar}\sum_{n=1}^{\infty} \frac{\mathcal{N}_n^2g_n(\omega)}{v^2} \mathcal{F}_{\lf}\left(t_n\right)\frac{1}{2\pi}\sum_{j=1}^{\infty}\sum_{k=1}^{\infty}\mathcal{Z}_{j,\lf}\mathcal{Z}_{k,\lf}\ee^{-\left(\alpha_{j,\lf}^2+\alpha_{k,\lf}^2\right)\frac{\BW^2}{8a^2}}I_{\lf}\left(\alpha_{j,\lf}\alpha_{j,\lf}\frac{\BW^2}{4a^2}\right),
\end{align}
where, in this case, we integrate over the whole space, exploiting the fact that the screened potential and the Gaussian beam go to zero far from the nanoparticle. The absolute dichroism in such situation becomes
\begin{align}
    \mathcal{AD}_0^{\lf}&=\frac{8e^2}{\hbar}\left[\mathcal{F}_{\lf}(t_n)-\mathcal{F_{-\lf}}(t_n)\right]\sum_{n=1}^{\infty} \frac{\mathcal{N}_n^2g_n(\omega)}{2\pi v^2} \sum_{j=1}^{\infty}\sum_{k=1}^{\infty}\mathcal{Z}_{j,\lf}\mathcal{Z}_{k,\lf}\ee^{-\left(\alpha_{j,\lf}^2+\alpha_{k,\lf}^2\right)\frac{\BW^2}{8a^2}}I_{\lf}\left(\alpha_{j,\lf}\alpha_{j,\lf}\frac{\BW^2}{4a^2}\right).
\end{align}

\subsubsection{Absolute dichroism without post-selection filter}
We can also use the analytical model to easily explore what will happen when the post-selection OAM filter is omitted. Energy-loss probability in such a case would be calculated as a sum of all possible final states $\lf$. Let us define the total EEL spectra for a VEB with initial topological charge $\li$ as
\begin{align}
    \nonumber
    \Gamma_{\li}(\omega) &= \sum_{\lf=-\infty}^{\infty} \Gamma_{\li}^{\lf}(\omega)\\
    &=\frac{8e^2}{\hbar}\sum_{\lf=-\infty}^{\infty}\sum_{n=1}^{\infty} \frac{\mathcal{N}_n^2g_n(\omega)}{v^2} \mathcal{F}_{\Delta\ell}\left(t_n\right) \frac{1}{\pi|\li|!}\left(\frac{2}{\BW^2}\right)^{|\li|+1}    \int_0^{\Rcut}R^{2|\li|+1} \mathrm{d}R\, \ee^{-\frac{2R^2}{\BW^2}} \left(\sum_{j=1}^{\infty} \mathcal{Z}_{j,\Delta\ell}J_{\Delta\ell}\left(\alpha_{j,\Delta\ell}\frac{R}{a}\right) \right)^2.
    \label{eq:Total_gamma}
\end{align}
Notice that in Eq.~\eqref{eq:Total_gamma} the initial topological charge $\li$ in explicit form is always in absolute value, so for $\Gamma_{-\li}$ we obtain same formula. However, the $\li$ is also present implicitly in the term $\Delta\ell$. When we sum over all final states $\lf$, we always find such a value of $\lf$ that matches the same $\Delta\ell$ in $\Gamma_{\li}$ spectra. This means that when the radial profile $f_{|\li|}(R)$ of the VEB is independent on the polarity of the topological charge, and we collect all the transmitted electrons on the detector (i.e. $\Qf\rightarrow\infty$), the absolute and relative dichroism yield zero. This result underlines the importance of the post-selection OAM filter in the experiment and emphasizes the dependence on the difference in OAM $\hbar\Delta\ell$ that enables the study of dichroism in EELS.

\subsubsection{Emergence of dichroism in non-chiral samples}
\begin{figure}[H]
    \centering
    \includegraphics[width=1\linewidth]{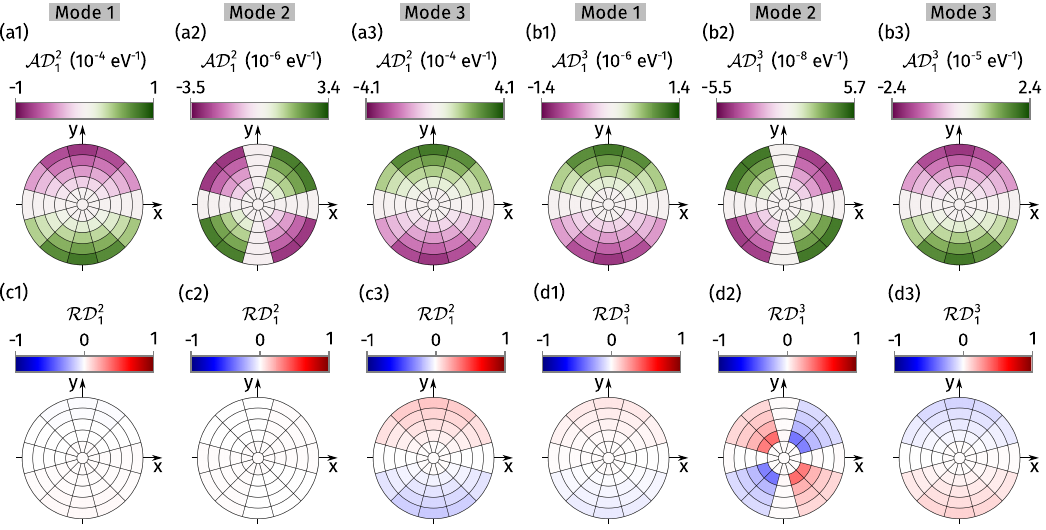}
    \caption{
    Extended calculations related to Fig.~4 of the main text: absolute and relative dichroism for an achiral thin silver ``C-shaped'' object with radius $R_0=50$~nm (pitch $d=0$~nm) exposed to a tilted beam. The tilt is defined in the same way as in Fig.~4(c). (a1--a3) show the dependence for the incident Laguerre-Gaussian beam with $\li=1$ and $\Delta\ell=1$. (b1--b3) Same as (a1--a3) for $\lf=3$ and $\Delta\ell=2$. The lower row shows relative dichroism for $\Delta\ell=1$ [(c1--c3)] and $\Delta\ell=2$ [(d1--d3)]. In all cases, the electron beam has parameters $\BW=10$~nm, energy 30~keV, and we consider a collection angle $\alphaf=5$~mrad.
    }
    \label{fig:SI_Fig6}
\end{figure}

In Fig.~\ref{fig:SI_Fig6}, we present results showing the dependence of absolute and relative dichroism for achiral ``C-shaped'' nanoparticle (i.e. thin silver single-twist helix with pitch $d=0$~nm) for a tilted Laguerre-Gaussian beam with the same initial topological charge $\li=1$ and different $\Delta\ell$ [$\Delta\ell=1$ in Fig.~\ref{fig:SI_Fig6}(a1--a3, c1--c3) and $\Delta\ell=2$ in Fig.~\ref{fig:SI_Fig6}(b1--b3, d1--d3)]. As we see, the achiral particle exhibits dichroism when the normal to the plane containing the ``C-shaped'' particle (in this case, the $z$-direction) is tilted with respect to the electron trajectory. However, thanks to the symmetry of surface charge, there are tilt directions where the dichroism is zero. When the electron's trajectory is in the $x$--$z$ plane (i.e. plane of surface charge symmetry) the dichroism is zero. For even modes, another plane emerges [see Fig.~\ref{fig:SI_Fig6}(a2, b2)]. For a larger change of OAM ($\Delta\ell=2$), we observe stronger relative dichroism for the second mode [Fig.~\ref{fig:SI_Fig6}(d2)], however, the absolute dichroism is in this case in the order of $10^{-8}$~eV, which is intensity comparable with experimental noise.

\subsubsection{Additional calculations}

The dependence of the relative dichroism defined by Eq.~(5) on the parameter $t_n$ is presented in Fig.~2. However, the energy of the plasmonic mode $\hbar\omega_n$ also depends on the geometry of the structure. It is therefore suitable to show the relative dichroism dependence in the same manner as was shown for absolute dichroism in Fig.~3. In Fig.~\ref{fig:SI_Fig1} we present dependence on the pitch and extended interval of electron energy for $\Delta\ell=1$, and $\Delta\ell=2$ for both absolute dichroism and relative dichroism. By lowering the electron energy, it is possible to probe in the regions of Fig.~2 where the parameter $t_n$ is up to 14.7, 26.6, and 36.3, respectively, for the first three modes. For helices with fixed geometry, the polarity of relative (and absolute) dichroism can significantly change with different acceleration voltages.

\newpage

\begin{figure}[H]
    \centering
    \includegraphics[width=1\linewidth]{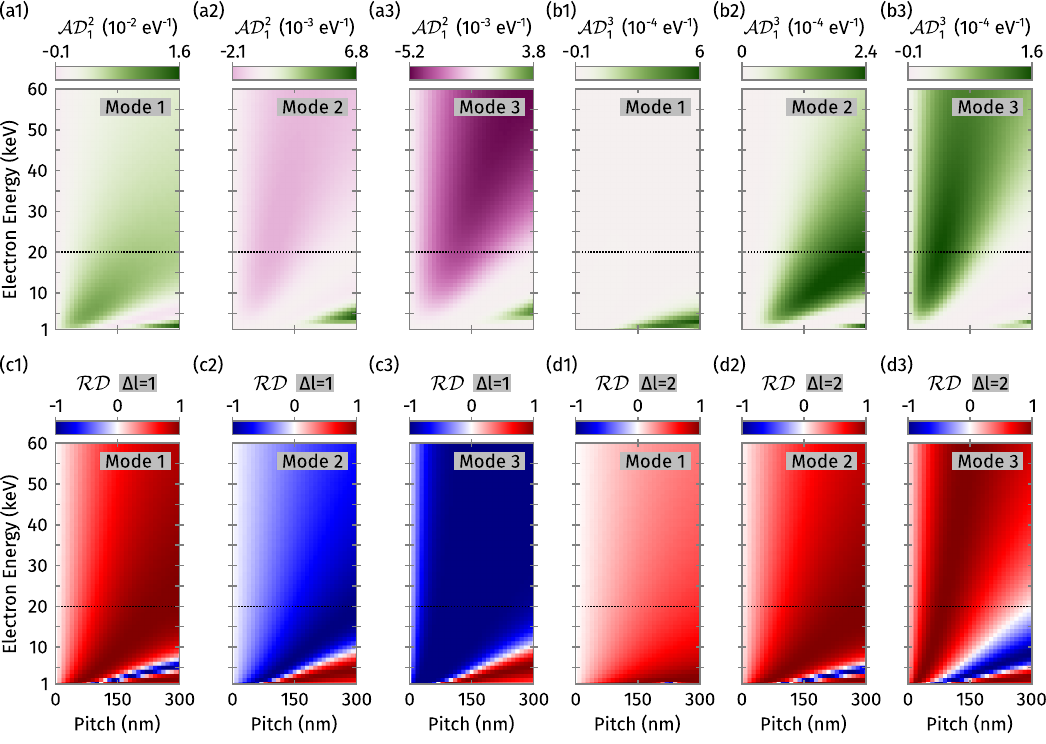}
    \caption{Extended calculations related to Fig.~3 of the main text: absolute (a1--b3) and relative (c1--d3) dichroism as a function of electron energy and pitch calculated for an infinitesimally thin silver single-twist helix ($R_0=50$~nm, a thickness of $\xi=10$~nm was used to determine the normalisation constant $\sigma_n$ from MNPBEM simulation). We considered the collection of all electrons, thus $\Qf\rightarrow\infty$. For all simulations of $\mathcal{AD}$, an electron beam with $\BW=10$~nm was used. Panels (a1--a3, c1--c3) show the dependence for the first three modes and the change in OAM is  $\Delta\ell=1$, while $\Delta\ell=2$ is considered in (b1--b3, d1--d3).
    Data above the dashed horizontal line describe the same region as in Fig.~3\mbox{(b1--d3)}. 
    The electron energy interval is extended to lower acceleration voltages (1\,keV--60\,keV), allowing us to explore higher values of the parameter $t_n$.}
    \label{fig:SI_Fig1}
\end{figure}

\begin{figure}[H]
    \centering
    \includegraphics[width=1\linewidth]{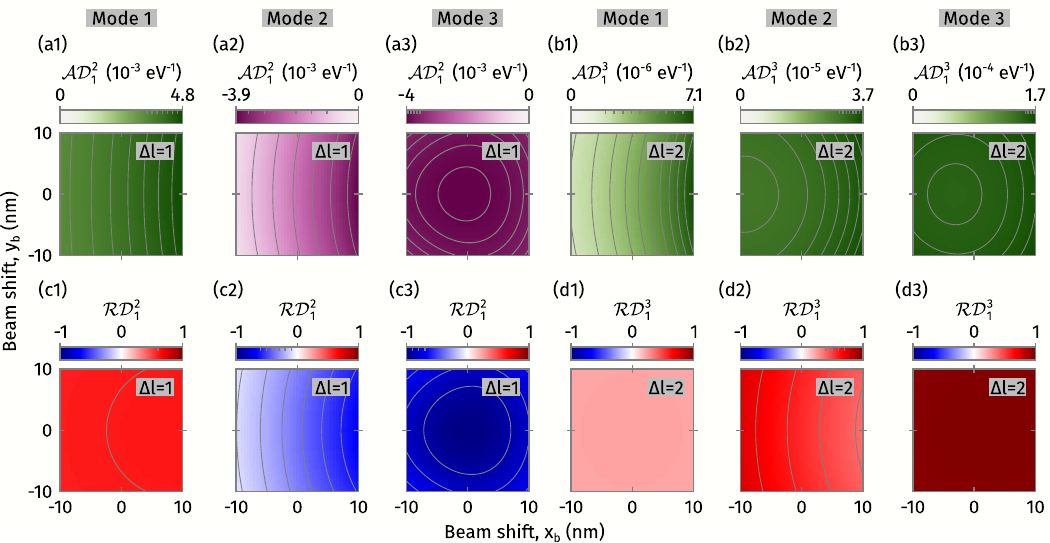}
    \caption{
    Extended calculations related to Fig.~4(c1--c3): absolute and relative dichroism for a thin silver single-twist helix with radius $R_0=50$~nm and pitch $d=100$\,nm exposed to a 30~keV Gaussian electron beam with a collection angle $\alphaf=5$~mrad. The electron beam's trajectory is parallel to the helix axis, impinging on the helix at position $\Rb_\mathrm{b}=(x_\mathrm{b}, y_\mathrm{b})$. The rectangle represents the same area as in Fig.~4(c1--c3). (a1--a3) show the dependence for the Laguerre-Gaussian beam with $\li=1$ and $\Delta\ell=1$ for the first three modes. (b1--b3) Same as (a1--a3) for $\lf=3$ and $\Delta\ell=2$. Ticks in colorbars mark the contour levels. The contour steps are $2\cdot10^{-4}$, $5\cdot10^{-4}$ and $1\cdot10^{-4}\,\mathrm{eV}^{-1}$ respectively in (a1--a3) and $5\cdot10^{-7}$, $1\cdot10^{-6}$ and $4\cdot10^{-6}\,\mathrm{eV}^{-1}$ respectively in (b1--b3). Relative dichroism calculations in the lower row were obtained for the same parameters as the above panels, but we note that these results are independent of exact $\li$ and $\lf$ and only depend on $\Delta \ell$. Contour step in (c1--d3) is 0.1, and only the visible contours are displayed in the colorbar.
    }
    \label{fig:SI_Fig3}
\end{figure}

\begin{figure}[H]
    \centering
    \includegraphics[width=1\linewidth]{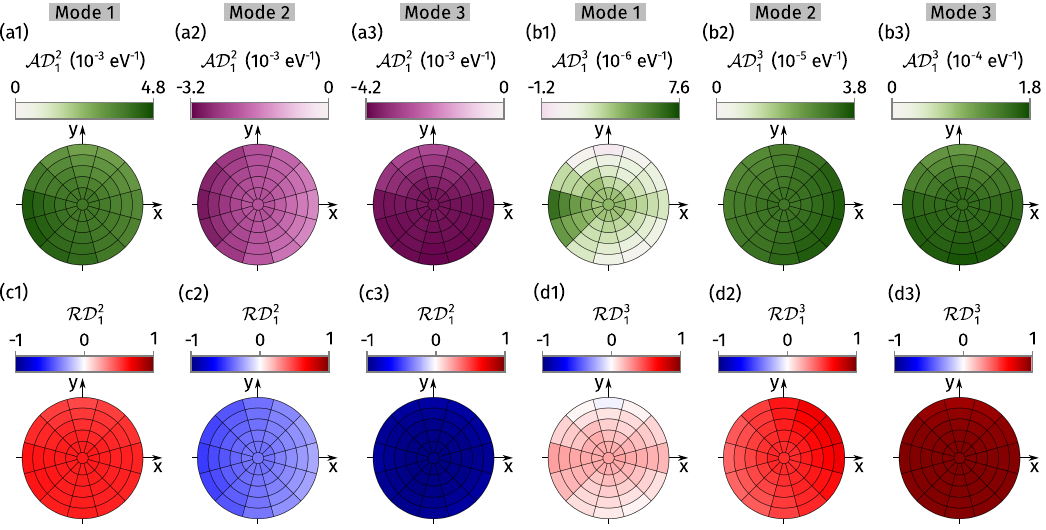}
    \caption{Extended calculations related to Fig.~4(d1--d3) of the main text: absolute and relative dichroism for a thin silver single-twist helix with radius $R_0=50$~nm and pitch $d=100$~nm exposed to a tilted beam, with the tilt and tilt steps defined the same way as in defined in Fig.~4(b). (a1--a3) $\mathcal{AD}_1^2$ for the first three modes. (b1--b3) Same as (a1--a3) for $\lf=3$ and $\Delta\ell=2$. The lower row shows relative dichroism corresponding to the parameters in the panels above. In all cases, the electron beam has parameters $\BW=10$~nm, energy 30~keV, and we consider a collection angle $\alphaf=5$~mrad. For the Gaussian beam with $\li=0$ as considered in Fig.~4(d1--d3), the absolute dichroism is scaled, and the relative dichroism for $\Delta\ell=\{1;2\}$ looks the same as in the (c1--d3).
    }
    \label{fig:SI_Fig2}
\end{figure}

\subsection{Details on boundary element method and numerical simulations in MNPBEM}
\label{Sec:MNPBEM}

\subsubsection{Formulation of boundary element method and electron energy-loss probability in the quasi-static regime and in SI units}
To establish expressions for the electron energy-loss probability in SI units, we followed the derivation from Ref.~\cite{garcia_de_abajo_numerical_1997}, making appropriate changes to Maxwell's equations and definitions of electrostatic potentials and Green's functions. Starting with Gauss's law of electrostatics relating electric displacement $\Db =  \varepsilon_{0} \varepsilon \Eb$ to external charge density $\rho_{\mathrm{ext}}$

\begin{align}
    \boldsymbol{\nabla} \cdot \Db & = \boldsymbol{\nabla} \cdot ( \varepsilon_{0} \varepsilon \Eb ) = - \boldsymbol{\nabla} \cdot \left( \varepsilon_{0} \varepsilon \boldsymbol{\nabla} \phi \right)  = \rho_{\mathrm{ext}},
\end{align}
where $\varepsilon_{0}$ is the vacuum permittivity, $\varepsilon$ is the relative permittivity, and the electrostatic potential $\phi$ has to satisfy Poisson's equation
    \begin{align}
    \nabla^2 \phi & = -\frac{\rho_{\mathrm{ext}}}{\varepsilon_{0} \varepsilon} - \frac{\boldsymbol{\nabla} \varepsilon \cdot \boldsymbol{\nabla} \phi}{\varepsilon},
\end{align}
where the second source term containing the permittivity gradient $ \boldsymbol{\nabla}  \varepsilon$ is related to inhomogeneities in dielectric response. The solution to the above equation can be conveniently expressed using the free-space Green's function

\begin{align}
    W_{0} (\rb,\rb^{\prime} ) = \frac{1}{4\pi \vert \rb -\rb^{\prime} \vert},
\end{align}
satisfying
\begin{align}
    \nabla^2 W_{0} (\rb,\rb^{\prime} ) = - \delta (\rb -\rb^{\prime}).
\end{align}
Denoting $\boldsymbol{\nabla}^{\prime}$ the differential operator that acts on primed coordinates, the electrostatic potential reads

\begin{align}
    \phi(\rb) & = \int\mathrm{d}^3\rb^{\prime}\,\, W_{0} (\rb,\rb^{\prime} ) \frac{\rho_{\mathrm{ext}} (\rb^{\prime}) }{ \varepsilon_{0} \varepsilon} + \int\mathrm{d}^3\rb^{\prime} \,\, W_{0} (\rb,\rb^{\prime} ) \frac{\boldsymbol{\nabla}^{\prime} \varepsilon(\rb^{\prime}) \cdot \boldsymbol{\nabla}^{\prime} \phi(\rb^{\prime}) }{\varepsilon} \nonumber \\
    & = \int\mathrm{d}^3\rb^{\prime} \frac{\rho_{\mathrm{ext}} (\rb^{\prime}) }{4 \pi \varepsilon_{0} \varepsilon \vert \rb - \rb^{\prime} \vert} + \int\mathrm{d}^3\rb^{\prime} \frac{\boldsymbol{\nabla}^{\prime} \varepsilon(\rb^{\prime}) \cdot \boldsymbol{\nabla}^{\prime} \phi(\rb^{\prime}) }{4 \pi \varepsilon \vert \rb - \rb^{\prime} \vert}.
\end{align}
Assuming the dielectric permittivity $\varepsilon$ is homogeneous everywhere except at abrupt interfaces between objects, the second source term can be expressed using the surface delta function $\delta_{\mathrm{s}}$ that defines an interface between two media with dielectric permittivities $\varepsilon_{1}$ and $\varepsilon_{2}$

\begin{align}
    \frac{\boldsymbol{\nabla} \varepsilon \cdot \boldsymbol{\nabla} \phi}{\varepsilon} = -\frac{\varepsilon_{0} \varepsilon \boldsymbol{\nabla} \phi}{\varepsilon_{0}} \cdot \boldsymbol{\nabla}\left( \frac{1}{\varepsilon} \right) = \frac{\Db}{\varepsilon_{0}} \cdot \boldsymbol{\nabla}\left( \frac{1}{\varepsilon} \right) = \frac{\Db}{\varepsilon_{0}} \cdot \nb_{\mathrm{s}} \delta_{\mathrm{s}} \left( \frac{1}{\varepsilon_{2}} - \frac{1}{\varepsilon_{1}} \right)\ \nonumber =  \left( \nb_{\mathrm{s}} \cdot \Db \right) \frac{\varepsilon_{1}-\varepsilon_{2}}{\varepsilon_{0} \varepsilon_{1} \varepsilon_{2}}  \delta_{\mathrm{s}},
\end{align}
where the surface normal $\nb_{\mathrm{s}}$ points from medium 1 to medium 2. Recognizing the expression in front of the surface delta function as an induced boundary charge density

\begin{align}
    \sigma (\ssb) = \left( \nb_{\mathrm{s}} \cdot \Db \right) \frac{\varepsilon_{1}-\varepsilon_{2}}{\varepsilon_{0} \varepsilon_{1} \varepsilon_{2}},
    \label{eq:NormDisplacement}
\end{align}
the electrostatic potential can be split into two contributions: one that represents the direct response to the external charge density

\begin{align}
    \phi_{\mathrm{ext}}(\rb) & = \int\mathrm{d}^3\rb^{\prime} \frac{\rho_{\mathrm{ext}} (\rb^{\prime}) }{4 \pi \varepsilon_{0} \varepsilon \vert \rb - \rb^{\prime} \vert},
\end{align}
and one that accounts for the presence of other objects through the boundary charge induced at interfaces

\begin{align}
     \phi_{\mathrm{bnd}}(\rb) & = \int\mathrm{d}^2\ssb \frac{\sigma (\ssb) }{4 \pi \vert \rb - \ssb \vert}.
     \label{eq:BndPot}
\end{align}
To find the boundary charge, we need to evaluate the normal component of the electric displacement in Eq.~\eqref{eq:NormDisplacement}. Note that there is ambiguity in choosing which side of the interface should be used for this evaluation---both are acceptable, the main difference resides in the treatment of the singularity lurking in Eq.~\eqref{eq:BndPot} when $\rb \rightarrow \ssb$

\begin{align}
    \lim_{t \rightarrow 0} \nb_{\mathrm{s}} \cdot \boldsymbol{\nabla} \frac{1}{\vert \ssb \pm t \nb_{\mathrm{s}} - \ssb^{\prime} \vert} = - \frac{ \nb_{\mathrm{s}} \cdot (\ssb - \ssb^{\prime})}{\vert \ssb - \ssb^{\prime} \vert^{3}} \mp 2\pi \delta (\ssb - \ssb^{\prime}) = F (\ssb , \ssb^{\prime}) \mp 2\pi \delta (\ssb - \ssb^{\prime}).
    \label{eq:Singularity1}
\end{align}

\noindent
The choice of the sign in front of the delta function depends on whether we approach the interface from the side of medium 1 (bottom sign) or medium 2 (top sign). Choosing the latter, the normal component of the electric displacement becomes

\begin{align}
    \nb_{\mathrm{s}} \cdot \Db (\ssb) & = \varepsilon_{0} \varepsilon_{2} \left\lbrack - \nb_{\mathrm{s}} \cdot \boldsymbol{\nabla} \phi_{\mathrm{ext}}(\ssb) - \nb_{\mathrm{s}} \cdot \int\mathrm{d}^2\ssb \frac{\sigma (\ssb^{\prime}) }{4 \pi } \boldsymbol{\nabla} \frac{1}{\vert \ssb - \ssb^{\prime} \vert}  \right\rbrack  \nonumber \\
    & = \varepsilon_{0} \varepsilon_{2} \left\lbrack - \nb_{\mathrm{s}} \cdot \boldsymbol{\nabla} \phi_{\mathrm{ext}}(\ssb) - \frac{1}{4\pi} \cdot \int\mathrm{d}^2\ssb \,\, F (\ssb , \ssb^{\prime}) \sigma (\ssb^{\prime}) + \frac{\sigma (\ssb)}{2} \right\rbrack = \frac{\varepsilon_{0} \varepsilon_{1} \varepsilon_{2}}{\varepsilon_{1}-\varepsilon_{2}} \sigma (\ssb).
\end{align}

\noindent
When inserted into Eq.~\eqref{eq:NormDisplacement}, we obtain a self-consistent equation for the boundary charge density

\begin{align}
    2 \pi \frac{\varepsilon_{2}+\varepsilon_{1}}{\varepsilon_{2}-\varepsilon_{1}} \sigma (\ssb) = \Lambda (\omega) \sigma (\ssb) = 4\pi \, \nb_{\mathrm{s}} \cdot \boldsymbol{\nabla} \phi_{\mathrm{ext}}(\ssb) + \int\mathrm{d}^2\ssb \,\, F (\ssb , \ssb^{\prime}) \sigma (\ssb^{\prime})
    \label{eq:SelfConsistentEq}
\end{align}
which, in the absence of an external stimulus, reduces to an eigenvalue equation with eigenvalues $\lambda_{n}$

\begin{align}
    2 \pi \lambda_{n} \sigma_{n} (\ssb) & = \int\mathrm{d}^2\ssb \,\, F (\ssb , \ssb^{\prime}) \sigma_{n} (\ssb^{\prime}),
    \label{eq:EigenvalEq}
\end{align}

\noindent
where the eigenvectors $\sigma_{n} (\ssb)$ form a complete basis set with the following orthogonality property

\begin{align}
    \int\mathrm{d}^2\ssb \int\mathrm{d}^2\ssb^{\prime} \frac{\sigma_{m} (\ssb) \sigma_{n}^{\ast} (\ssb^{\prime})}{\vert \ssb - \ssb^{\prime} \vert} & = \delta_{mn}.
    \label{eq:Orthogonality}
\end{align}

\noindent
This allows us, ultimately, to expand the inhomogeneous term in Eq.~\eqref{eq:SelfConsistentEq} in this basis set and express the boundary charge as a linear combination of the eigenvectors (or in other words, the eigenmodes of the studied nanoparticle)

\begin{align}
    \sigma (\ssb) & = \sum_{n,\mu} \frac{f_{n\mu} }{\Lambda - 2\pi \lambda_{n} } \sigma_{n} (\ssb), \\
    f_{n\mu} & = 4\pi \int\mathrm{d}^2\ssb \int\mathrm{d}^2\ssb^{\prime} \frac{\nb_{\mathrm{s}} \cdot \boldsymbol{\nabla} \phi_{\mathrm{ext},\mu}(\ssb)}{\vert \ssb - \ssb^{\prime} \vert} \sigma_{n}^{\ast} (\ssb^{\prime}),
    \label{eq:fnmu}
\end{align}

\noindent
with the summation over the index $\mu$ allowing for the general case when the external charge density extends over multiple media (e.g. an electron beam passing through a particle).

Evaluation of EEL spectra in this work relies heavily on the introduction of a screened potential $W_{\mathrm{SI}} (\rb , \rb^{\prime},\omega)$ and its expansion in terms of potentials $w_{n}(\rb)$ associated with individual eigenmodes of the helix nanoparticle

\begin{align}
    W_{\mathrm{SI}} (\rb , \rb^{\prime}, \omega) = \sum_{n} h_{n} (\omega) \, w_{n}(\rb) w_{n}^{\ast} (\rb^{\prime}) = \sum_{n} h_{n} (\omega) \int\mathrm{d}^2\ssb \frac{\sigma_{n} (\ssb) }{4 \pi \vert \rb - \ssb \vert} \int\mathrm{d}^2\ssb^{\prime} \frac{\sigma_{n}^{\ast} (\ssb^{\prime}) }{4 \pi \vert \rb^{\prime} - \ssb^{\prime} \vert}.
    \label{eq:WExpanded}
\end{align}

\noindent
The presence of the subscript SI indicates that the above screened potential is defined in SI units and is, therefore, not exactly the same as the one appearing in the main text.

To find the expression for the spectral function $h_{n} (\omega)$ that determines the weight of the $n$-th eigenmode in the above expansion, one needs to evaluate the coefficients $f_{n\mu}$ in Eq.~\eqref{eq:fnmu} when the external potential is set to

\begin{align}
    \phi_{\mathrm{ext},\mu}(\ssb, \rb^{\prime}) & = \frac{1}{4 \pi \varepsilon_{0} \varepsilon_{\mu} \vert \ssb - \rb^{\prime} \vert}.
\end{align}

\noindent
Careful treatment of the singularity that appears in the external potential when $\rb^{\prime} \rightarrow \ssb$ yields 

\begin{align}
    \nb_{\mathrm{s}} \cdot \boldsymbol{\nabla} \phi_{\mathrm{ext},\mu}(\ssb, \rb^{\prime}) =\frac{1}{4 \pi \varepsilon_{0} \varepsilon_{\mu} }  \lim_{t \rightarrow 0} \nb_{\mathrm{s}} \cdot \boldsymbol{\nabla} \frac{1}{\vert \ssb \mp t \nb_{\mathrm{s}} - \rb^{\prime} \vert} = \frac{F (\ssb , \rb^{\prime}) \pm 2\pi \delta (\ssb - \rb^{\prime})}{4 \pi \varepsilon_{0} \varepsilon_{\mu}}.
    \label{eq:Singularity2}
\end{align}

\noindent
Note that, unlike in Eq.~\eqref{eq:Singularity1}, we are approaching with the source and not with the observation point, which leads to a reversal in the signs.

For the sake of simplicity, let us now limit ourselves to the case where both the source and the observation point lie in medium 2 so that only the coefficients $f_{n2}$ are non-zero and the upper sign in Eq.~\eqref{eq:Singularity1} is considered. The screened potential then becomes

\begin{align}
    W_{\mathrm{SI}} (\rb , \rb^{\prime}, \omega) & = \int\mathrm{d}^2\ssb \frac{\sigma (\ssb) }{4 \pi \vert \rb - \ssb \vert} = \int\mathrm{d}^2\ssb \sum_{n} \frac{f_{n2} }{\Lambda(\omega) - 2\pi \lambda_{n} } \sigma_{n} (\ssb) \nonumber \\
    & = \sum_{n} \frac{1}{\Lambda(\omega) - 2\pi \lambda_{n}} \int\mathrm{d}^2\ssb \frac{\sigma_{n} (\ssb) }{4 \pi \vert \rb - \ssb \vert} \,\frac{1}{\varepsilon_{0} \varepsilon_{2}} \int\mathrm{d}^2\ssb^{\prime} \int\mathrm{d}^2\ssb^{\prime\prime} \frac{F ( \ssb^{\prime}, \rb^{\prime} ) + 2\pi \delta (\ssb^{\prime} - \rb^{\prime})}{\vert \ssb^{\prime} - \ssb^{\prime\prime} \vert } \sigma_{n}^{\ast} (\ssb^{\prime\prime}).
\end{align}

\noindent
Next, we evaluate the potential generated by an eigenmode $\sigma_{m}$ at a position $\rb$ using the above expression for the screened potential

\begin{align}
    \int\mathrm{d}^2\rb^{\prime} & \,\, W_{\mathrm{SI}} (\rb , \rb^{\prime}, \omega) \, \sigma_{m} (\rb^{\prime}) \nonumber \\
    & =\frac{1}{\varepsilon_{0} \varepsilon_{2}} \sum_{n} \frac{1}{\Lambda(\omega) - 2\pi \lambda_{n}} \int\mathrm{d}^2\ssb \frac{\sigma_{n} (\ssb) }{4 \pi \vert \rb - \ssb \vert} \int\mathrm{d}^2\ssb^{\prime} \int\mathrm{d}^2\ssb^{\prime\prime} \frac{\sigma_{n}^{\ast} (\ssb^{\prime\prime})}{\vert \ssb^{\prime} - \ssb^{\prime\prime} \vert } \int\mathrm{d}^2\rb^{\prime} \left\lbrack F ( \ssb^{\prime}, \rb^{\prime} ) + 2\pi \delta (\ssb^{\prime} - \rb^{\prime}) \right\rbrack \sigma_{m} (\rb^{\prime}) \nonumber \\
    & = \frac{1}{\varepsilon_{0} \varepsilon_{2}} \sum_{n} \frac{1}{\Lambda(\omega) - 2\pi \lambda_{n}} \int\mathrm{d}^2\ssb \frac{\sigma_{n} (\ssb) }{4 \pi \vert \rb - \ssb \vert} \int\mathrm{d}^2\ssb^{\prime} \int\mathrm{d}^2\ssb^{\prime\prime} \frac{\sigma_{n}^{\ast} (\ssb^{\prime\prime})}{\vert \ssb^{\prime} - \ssb^{\prime\prime} \vert } \left\lbrack 2\pi \lambda_{n} \sigma_{m} (\ssb^{\prime}) + 2\pi \sigma_{m} (\ssb^{\prime}) \right\rbrack \nonumber \\
    & = \frac{2\pi}{\varepsilon_{0} \varepsilon_{2}} \sum_{n} \frac{\lambda_{n} +1}{\Lambda(\omega) - 2\pi \lambda_{n}} \int\mathrm{d}^2\ssb \frac{\sigma_{n} (\ssb) }{4 \pi \vert \rb - \ssb \vert} \int\mathrm{d}^2\ssb^{\prime} \int\mathrm{d}^2\ssb^{\prime\prime} \frac{\sigma_{m} (\ssb^{\prime}) \sigma_{n}^{\ast} (\ssb^{\prime\prime})}{\vert \ssb^{\prime} - \ssb^{\prime\prime} \vert } \nonumber \\
    & = \frac{2\pi}{\varepsilon_{0} \varepsilon_{2}} \sum_{n} \frac{\lambda_{n} +1}{\Lambda(\omega) - 2\pi \lambda_{n}} \int\mathrm{d}^2\ssb \frac{\sigma_{n} (\ssb) }{4 \pi \vert \rb - \ssb \vert} \delta_{mn} = \frac{2\pi}{\varepsilon_{0} \varepsilon_{2}} \frac{\lambda_{m} +1}{\Lambda(\omega) - 2\pi \lambda_{m}} \int\mathrm{d}^2\ssb \frac{\sigma_{m} (\ssb) }{4 \pi \vert \rb - \ssb \vert},
    \label{eq:Overlap1}
\end{align}
where we subsequently exploited Eq.~\eqref{eq:EigenvalEq} (eigenvalue equation) and the orthogonality property given by Eq.~\eqref{eq:Orthogonality}. If we now replace the screened potential with its expansion from Eq.~\eqref{eq:WExpanded} and evaluate the above integral again, we arrive at

\begin{align}
    \int\mathrm{d}^2\rb^{\prime}  \,\, W_{\mathrm{SI}} (\rb , \rb^{\prime}, \omega) \, \sigma_{m} (\rb^{\prime}) &= \sum_{n} h_{n} (\omega) \int\mathrm{d}^2\ssb \frac{\sigma_{n} (\ssb) }{4 \pi \vert \rb - \ssb \vert} \int\mathrm{d}^2\rb^{\prime} \int\mathrm{d}^2\ssb^{\prime} \frac{\sigma_{m} (\rb^{\prime}) \sigma_{n}^{\ast} (\ssb^{\prime}) }{4 \pi \vert \rb^{\prime} - \ssb^{\prime} \vert} \nonumber \\
    & = \sum_{n} h_{n} (\omega) \int\mathrm{d}^2\ssb \frac{\sigma_{n} (\ssb) }{4 \pi \vert \rb - \ssb \vert} \frac{\delta_{mn} }{4\pi} = \frac{h_{m} (\omega)}{4\pi} \int\mathrm{d}^2\ssb \frac{\sigma_{m} (\ssb) }{4 \pi \vert \rb - \ssb \vert}.
\end{align}

\noindent
The spectral function $h_{n} (\omega)$ is then readily obtained by comparing the term in front of the final integral with its counterpart in Eq.~\eqref{eq:Overlap1}

\begin{align}
    h_{n} (\omega) = \frac{8\pi^2}{\varepsilon_{0} \varepsilon_{2}} \frac{\lambda_{n} +1}{\Lambda(\omega) - 2\pi \lambda_{n}} = \frac{4\pi}{\varepsilon_{0} \varepsilon_{2}} \frac{ (\lambda_{n} +1 ) (\varepsilon_{2} - \varepsilon_{1} ) }{(\varepsilon_{2} + \varepsilon_{1} ) - (\varepsilon_{2} - \varepsilon_{1} ) \lambda_{n} } = - \frac{4\pi}{\varepsilon_{0} \varepsilon_{2}} + \frac{4\pi}{\varepsilon_{0} } \frac{2}{\varepsilon_{1} (1+\lambda_{n}) + \varepsilon_{2} (1-\lambda_{n}) }.
\end{align}

\noindent
Interestingly, the only frequency-dependent terms entering the spectral function are dielectric functions, eigenvalues $\lambda_{n}$ are dictated by geometry, and they are frequency-independent (as expected in non-retarded approximation).

The final step towards the sought expression for electron energy-loss probability is to find its link to the screened potential. According to Poynting's theorem, the rate at which an electron beam loses energy due to an interaction with a field induced by the beam's surroundings reads

\begin{align}
    \frac{\mathrm{d} E (\rb,t )}{\mathrm{d} t } = - \jb (\rb,t )\cdot \Eb_{\mathrm{ind}} (\rb,t ) = \jb (\rb,t )\cdot \boldsymbol{\nabla} \phi_{\mathrm{ind}} (\rb,t ) = \boldsymbol{\nabla} \cdot ( \phi_{\mathrm{ind}} \, \jb ) - \phi_{\mathrm{ind}} ( \boldsymbol{\nabla} \cdot \jb ) = \boldsymbol{\nabla} \cdot ( \phi_{\mathrm{ind}} \, \jb ) + \phi_{\mathrm{ind}} \frac{\partial \rho_{\mathrm{ext}} }{\partial t},
    \label{eq:LocalDissipationRate}
\end{align}
where we used the fact that the charge and current densities are tied through the continuity equation $\boldsymbol{\nabla} \cdot \jb + \frac{\partial \rho_{\mathrm{ext}} }{\partial t} = 0$. The global dissipation rate at any given time is then obtained by integration over the entire space

\begin{align}
    \frac{\mathrm{d} E ( t )}{\mathrm{d} t } = \int_V \mathrm{d}^3\rb \frac{\mathrm{d} E (\rb,t )}{\mathrm{d} t } = \oint_{\partial V} \mathrm{d} \Ab  \cdot \big\lbrack \phi_{\mathrm{ind}} (\rb,t ) \, \jb (\rb,t ) \big\rbrack + \int_V \mathrm{d}^3\rb \,\, \phi_{\mathrm{ind}} (\rb,t ) \frac{\partial \rho_{\mathrm{ext}} (\rb,t ) }{\partial t}.
\end{align}

\noindent
Utilizing the divergence theorem, the volume integration of the first term in Eq.~\eqref{eq:LocalDissipationRate} results in integration over the space boundary. Assuming this boundary is located sufficiently far away from any objects, the induced potential will vanish there and only the second term will contribute to the overall loss rate. The total energy loss is then obtained by integrating the above dissipation rate over time

\samepage{\begin{align}
    \Delta E & = \!\!\!\int\limits_{-\infty}^{\infty} \mathrm{d}t \frac{\mathrm{d} E ( t )}{\mathrm{d} t } = \int_V \mathrm{d}^3\rb \!\! \int\limits_{-\infty}^{\infty} \mathrm{d}t \,\, \phi_{\mathrm{ind}} (\rb,t ) \frac{\partial \rho_{\mathrm{ext}} (\rb,t ) }{\partial t} \nonumber \\
    & = \int_V \mathrm{d}^3\rb \Bigg\lbrace \Big\lbrack \rho_{\mathrm{ext}} (\rb,t ) \,\phi_{\mathrm{ind}} (\rb,t )  \Big\rbrack_{-\infty}^{\infty} \!\! - \!\! \int\limits_{-\infty}^{\infty} \mathrm{d}t \,\, \rho_{\mathrm{ext}} (\rb,t ) \frac{\partial \phi_{\mathrm{ind}} (\rb,t ) }{\partial t} \Bigg\rbrace = - \int_V \mathrm{d}^3\rb \!\! \int\limits_{-\infty}^{\infty} \mathrm{d}t \,\, \rho_{\mathrm{ext}} (\rb,t ) \frac{\partial \phi_{\mathrm{ind}} (\rb,t ) }{\partial t},
\end{align}}

\noindent
where we employed integration by parts and the fact that both at the beginning and at the end, the electrons forming the beam are far away from the probed objects, and the induced potential is essentially zero. Noting that the electrostatic potential is a real quantity and its Fourier components have to satisfy the condition $\phi_{\mathrm{ind}} (\rb,-\omega ) = \phi_{\mathrm{ind}}^{\ast} (\rb,\omega )$, we can replace $\phi_{\mathrm{ind}} (\rb,t )$ in the above equation by its Fourier spectrum in the form

\begin{align}
    \phi_{\mathrm{ind}} (\rb,t ) = \!\!\!\int\limits_{-\infty}^{\infty} \mathrm{d}\omega \,\, \phi_{\mathrm{ind}} (\rb,\omega ) \,\, \ee^{- i \omega t} = \int\limits_{0}^{\infty} \mathrm{d}\omega \,\, 2 \, \Ree \Big\lbrace \phi_{\mathrm{ind}} (\rb,\omega ) \,\, \ee^{- i \omega t} \Big\rbrace,
\end{align}
and the energy loss becomes

\begin{align}
    \Delta E & = - \int_V \mathrm{d}^3\rb \!\!\! \int\limits_{-\infty}^{\infty} \mathrm{d}t \,\, \rho_{\mathrm{ext}} (\rb,t ) \int\limits_{0}^{\infty} \mathrm{d}\omega \,\, 2 \Imm \Big\lbrace \omega \, \phi_{\mathrm{ind}} (\rb,\omega ) \,\, \ee^{- i \omega t} \Big\rbrace \nonumber \\
    & = \int\limits_{0}^{\infty} \mathrm{d}\omega \,\,4\pi\omega \int_V \mathrm{d}^3\rb \,\, \Imm \Big\lbrace - \phi_{\mathrm{ind}} (\rb,\omega ) \,\, \frac{1}{2\pi} \!\! \int\limits_{-\infty}^{\infty} \mathrm{d}t \,\, \rho_{\mathrm{ext}} (\rb,t ) \, \ee^{- i \omega t} \Big\rbrace = \int\limits_{0}^{\infty} \mathrm{d}\omega \,\,4\pi\omega \int_V \mathrm{d}^3\rb \,\, \Imm \Big\lbrace - \rho_{\mathrm{ext}}^\ast (\rb,\omega ) \, \phi_{\mathrm{ind}} (\rb,\omega ) \Big\rbrace \nonumber \\
    &  = \int\limits_{0}^{\infty} \mathrm{d}( \hbar \omega )\,\frac{\hbar \omega}{e}\, \frac{4\pi e}{\hbar^{2}} \int_V \mathrm{d}^3\rb \int_V \mathrm{d}^3\rb^{\prime} \,\, \Imm \Big\lbrace - \rho_{\mathrm{ext}}^\ast (\rb,\omega ) \,W_{\mathrm{SI}} (\rb, \rb^{\prime} ,\omega ) \, \rho_{\mathrm{ext}} (\rb^{\prime},\omega ) \Big\rbrace = \int\limits_{0}^{\infty} \mathrm{d}( \hbar \omega )\,\frac{\hbar \omega}{e} \, \Gamma(\omega),
    \label{eq:EnergyLossGen}
\end{align}
where we expressed the induced potential using the screened potential introduced earlier. The particular choice of the physical constants in front of the electron energy-loss probability function $\Gamma(\omega)$ ensures that its physical dimension is $\mathrm{eV}^{-1}$ and the energy loss is in Joules.

To clearly identify all the changes brought by the transfer from atomic to SI units, it is convenient to focus on a specific case that can also be easily tested numerically by an established tool such as MNPBEM. To that end, we chose to evaluate the loss probability function $\Gamma(\omega)$ for a single electron propagating along a non-penetrating trajectory parallel to the $z$-axis with a speed $v$ at a lateral distance $\rb_{\mathrm{b} }$ from the origin. The charge density then takes the form

\begin{align}
    \rho_{\mathrm{ext}} (\rb,\omega ) = \frac{1}{2\pi} \!\!\!\int\limits_{-\infty}^{\infty} \mathrm{d}t \,\, \rho_{\mathrm{ext}} (\rb,t ) \,\, \ee^{i \omega t} = \frac{-e}{2\pi} \!\!\!\int\limits_{-\infty}^{\infty} \mathrm{d}t \,\, \delta(\Rb - \Rb_{\mathrm{b} }) \, \delta (z-vt ) \,\, \ee^{i \omega t} = - \frac{e}{2 \pi v} \delta(\Rb - \Rb_{\mathrm{b} }) \, \ee^{i \omega z /v},
\end{align}
and the energy loss reads

\begin{align}
    \Gamma(\omega) & = \frac{e^3}{\pi \hbar^{2} v^{2} } \!\!\int\limits_{-\infty}^{\infty} \! \mathrm{d}{z} \!\! \int\limits_{-\infty}^{\infty} \! \mathrm{d}{z}^{\prime} \,\, \Imm \Big\lbrace - W_{\mathrm{SI}} (\Rb_{\mathrm{b} },z, \Rb_{\mathrm{b} },z^{\prime},\omega ) \, \ee^{ - i \omega (z-z^{\prime})/v } \Big\rbrace \nonumber \\
    & = \frac{e^3}{\pi \hbar^{2} v^{2} }\!\!\int\limits_{-\infty}^{\infty} \! \mathrm{d}{z} \!\! \int\limits_{-\infty}^{\infty} \! \mathrm{d}{z}^{\prime} \,\, \Imm \bigg\lbrace - \sum\limits_{n} h_{n} (\omega) \, w_{n}(\Rb_{\mathrm{b} },z,\omega) \, \ee^{ - i \omega z/v } \, w_{n}^{\ast} (\Rb_{\mathrm{b} },z^{\prime},\omega) \, \ee^{ i \omega z^{\prime}/v } \bigg\rbrace \nonumber \\
    & = \frac{e^3}{\pi \hbar^{2} v^{2} } \, \sum\limits_{n} \Imm \big\lbrace - h_{n} (\omega)  \big\rbrace \, \left\vert \int\limits_{-\infty}^{\infty} \! \mathrm{d}{z} \,\,  w_{n}(\Rb_{\mathrm{b} },z,\omega) \, \ee^{ - i \omega z/v } \right\vert^{2}.
\end{align}

\noindent
Apparently, volume integrals in Eq.~\eqref{eq:EnergyLossGen} are now reduced to line integrals along the $z$-coordinate that can be worked out analytically. Introducing projected potential $w_{n}(\Rb,\omega)$

\begin{align}
 w_{n}(\Rb,\omega) = \!\!\!\int\limits_{-\infty}^{\infty} \! \mathrm{d}z \,\, w_{n}(\Rb,z) \, \ee^{ - i \omega z/v } = \frac{1}{2\pi} \int \mathrm{d}^2\ssb \,\,\sigma_n(\ssb) \, \ee^{-\ii\omega s_z/v}K_0\left(\frac{\omega\lvert\Rb-\ssb_\perp\rvert}{v}\right),
 \label{Eq:wnProjectedSI}
\end{align}

\noindent
where $\ssb_{\perp} = (s_x,s_y)$ are lateral coordinates of the surface element $\mathrm{d}^2\ssb $, and further assuming that the electron probes a single nanoparticle with permittivity $\varepsilon_{1}=\varepsilon (\omega)$ embedded in vacuum ($\varepsilon_{2}=1$), the electron energy-loss probability function turns out to be

\begin{align}
    \Gamma(\omega) = \frac{e^3}{\pi \hbar^{2} v^{2} } \, \sum\limits_{n} \, \Imm \big\lbrace - h_n (\omega) \big\rbrace \, \vert w_{n}(\Rb_{\mathrm{b} },\omega) \vert^{2},
    \label{Eq:GammaSI}
\end{align}

\noindent
with the spectral function
\begin{align}
    h_n(\omega) = \frac{4\pi}{\varepsilon_{0} } \frac{2}{\varepsilon (\omega) (1+\lambda_{n}) + (1-\lambda_{n}) }.
\end{align}

\noindent
In our calculations, we assume silver as the material of the helix. To describe the optical properties of silver, we use the Drude model, where the dielectric function is defined as

\begin{align}
    \varepsilon(\omega) = \varepsilon_\infty - \frac{\wp^2}{\omega^2 + \ii \omega \gamma_\mathrm{d}},
\end{align}
where $\varepsilon_\infty=4$ is the background permittivity, $\wp=9.17$~eV is the plasma frequency, and $\gamma_\mathrm{d}=0.021$~eV characterizes the damping.

Note that the spectral function $g_{n}(\omega)$ appearing in the expansion of the screened potential $W (\rb, \rb^{\prime},\omega )$ in the main text is not the same as $h_n(\omega)$. To establish the link between the two and ensure that the energy-loss probability function is consistently expressed in $\mathrm{eV}^{-1}$ throughout the paper, we evaluated Eq.~\eqref{Eq:GammaNR2} for the specific case of a tightly focused electron beam. In that scenario, the incident wave can be expressed as
\begin{align}
    \left\lvert \psi_\mathrm{\perp,i}(\Rb) \right\rvert^2=\delta(\Rb - \Rb_\mathrm{b}),
    \label{Eq:delta-function_beam}
\end{align}
 where $\Rb_\mathrm{b}$ is the electron beam position. We consider an infinite detector and final states in the form of plane waves
\begin{align}
    \psi_\mathrm{\perp, f}(\Rb) = \frac{1}{\sqrt{A}}\ee^{\ii\Qb\cdot\Rb}.
\end{align}
After insertion into Eq.~\eqref{Eq:GammaNR2}, we obtain
\begin{align}
    \Gamma^\mathrm{NR}(\omega)&=\frac{e^2}{\pi\hbar v^2}\frac{1}{(2\pi)^2} \sum_{n=1}^{\infty} g_n(\omega) \int\mathrm{d}^2\Qb \int\mathrm{d}^2\Rb \int\mathrm{d}^2\Rb' \,\psi^\ast_\mathrm{\perp,i}(\Rb)\psi_\mathrm{\perp,i}(\Rb')\,\ee^{\ii\Qb\cdot(\Rb-\Rb')} \,w_n(\Rb,\omega)w_n^\ast(\Rb',\omega)
    \nonumber \\
    &=\frac{e^2}{\pi\hbar v^2}\sum_{n=1}^{\infty} g_n(\omega) \int\mathrm{d}^2\Rb\, \left\lvert\psi_\mathrm{\perp,i}(\Rb)\right\rvert^2\, \left\lvert w_n(\Rb,\omega)\right\rvert^2
    \nonumber =\frac{e^2}{\pi\hbar v^2}\sum_{n=1}^{\infty} g_n(\omega) \int\mathrm{d}^2\Rb\, \delta(\Rb - \Rb_\mathrm{b})\, \left\lvert w_n(\Rb,\omega)\right\rvert^2
    \nonumber \\
    &=\frac{e^2}{\pi\hbar v^2}\sum_{n=1}^{\infty} g_n(\omega) \, \left\lvert w_n(\Rb_\mathrm{b},\omega)\right\rvert^2.
\end{align}

\noindent
Clearly, the above energy-loss probability function can match the one given by Eq.~\eqref{Eq:GammaSI} only when 

\begin{align}
    g_n(\omega) = \frac{e}{\hbar} \Imm \big\lbrace - h_n(\omega) \big\rbrace = \frac{4\pi e}{\varepsilon_{0} \hbar } \, \Imm \bigg\lbrace \frac{-2}{\varepsilon (\omega) (1+\lambda_{n}) + (1-\lambda_{n}) } \bigg\rbrace.
    \label{Eq:gnSpectralFun}
\end{align}

\subsubsection{Normalization of helix eigenmodes}
\label{Sec:EigenmodeNormalization}
For the helix nanoparticles presented in this paper, one can either calculate the eigenmodes $\sigma_{n} (\ssb)$ numerically, for example, using the eigensolver implemented in the MNPBEM toolbox, or set up analytical Ansatz eigenmodes with an ideal cosine dependence. We should note that in both cases, the eigenmodes need to be normalized first to satisfy the orthogonality condition given by Eq.~\eqref{eq:Orthogonality}. Denoting $\sigma_n^{\prime} (\ssb)$ the original eigenmode, the norm $\sigma_n = \sigma_n^{\prime} (\ssb) / \sigma_n (\ssb)$ then reads

\begin{align}
    \sigma_n = \sqrt{ \int\mathrm{d}^2\ssb \int\mathrm{d}^2\ssb^{\prime} \frac{\sigma_{n}^{\prime} (\ssb) \, \sigma_{n}^{\prime\ast} (\ssb^{\prime})}{\vert \ssb - \ssb^{\prime} \vert} }. 
    \label{eq:EigenNorm}
\end{align}

\noindent
Since the analytical evaluation of the norm proved to be tricky, we mapped the Ansatz eigenmode distribution onto a discretized helix mesh generated in MNPBEM and calculated the norm using the same procedure as for the numerical eigenmodes. This procedure exploits the fact that the surface charge density can be approximately treated as a slowly varying function that is constant within a single discretization element $\Delta S_{i}$ and the integration subsequently turns into a summation over the discrete elements

\begin{align}
    \sigma_n = \sqrt{ \sum_i \sum_{j} \Delta S_{i} \, \sigma_{n}^{\prime}(\ssb_i) \, G (\ssb_i,\ssb_j ) \, \sigma_{n}^{\prime\ast} (\ssb_j) },
\end{align}

\noindent
where elements of the Green's function matrix

\begin{align}
    G (\ssb_i,\ssb_j ) = \int_{\Delta S_j} \!\! \mathrm{d}^2\ssb^{\prime} \frac{1}{\vert \ssb_i - \ssb^{\prime} \vert }
\end{align}
are conveniently computed by the MNPBEM solver and can be easily accessed.

\subsubsection{Validity of the non-retarded analytical approximation and effects of retardation on mode energy}

So far in this work, we used the analytically calculated screened potential in all simulations except Fig.~\ref{fig:Fig1}. However, we should also discuss the accuracy of this approximate approach. In this section, we present the comparison of the potentials calculated in the MNPBEM for a thick silver single-twist helix with the charge distribution calculated by solving Maxwell's equations and an infinitesimally thin silver single-twist helix with a charge distribution described by cosine dependence. In Fig.~\ref{fig:SI_Fig4}, we present the absolute value and phase of the functions $w_n$ for the first free modes, and we compare them with the MNPBEM simulation. The accuracy of the approximation is outstanding. The first mode exhibits very good accuracy in the absolute value (a2) and phase (a4), and is even better in the area denoted by circles, which represents the region where 95\% of Laguerre-Gaussian beam intensity lies. The calculations for the second and third modes are also very accurate. The large values [around 80\% in sub-figures (b2) and (c2)] are caused by dividing by a very small number. Outside these areas, the analytical result is similar to the one from MNPBEM. The phase of the second and third mode potential exhibits a point where the phase has a singularity (b3) and (c3). When we subtract the phase of potential calculated by MNPBEM from the analytical one, we see the singularity here as well. This singularity in phase does not cause any problems in the calculations, because the absolute value of $w_n$ is, in this case, zero. In the sub-figures (b4) and (c4), one can see that the difference in phases is close to zero outside the singular points.

\begin{figure}
    \centering
    \includegraphics[width=1\linewidth]{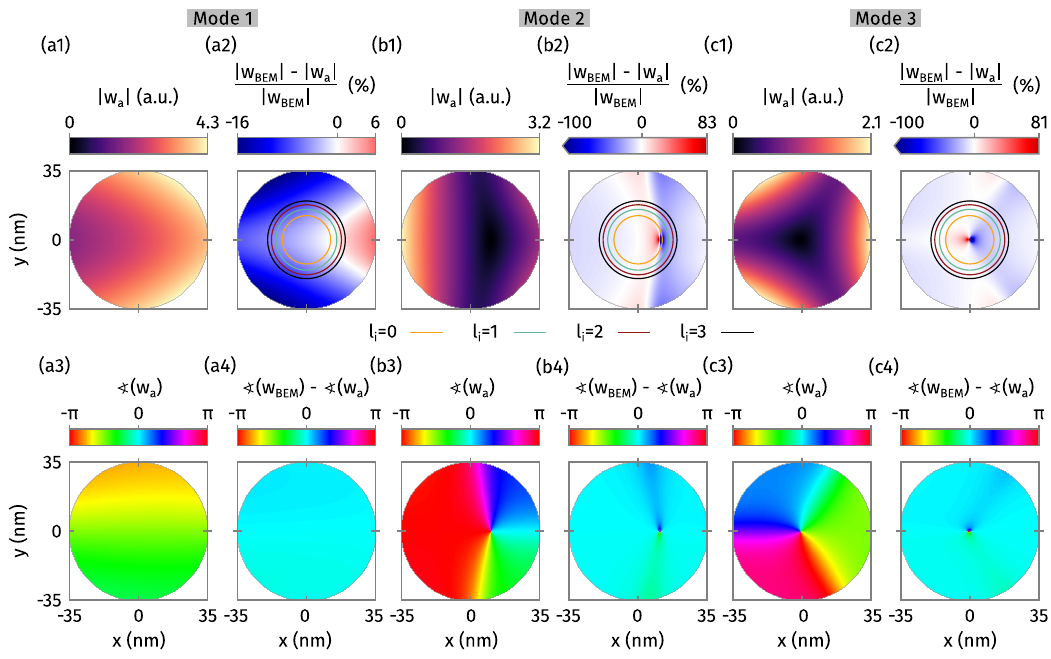}
    \caption{
    Comparison of the function $w$ for an infinitesimal thin silver single-twist helix calculated analytically ($w_\mathrm{a}$) and for a thick silver single-twist helix calculated numerically by MNPBEM ($w_\mathrm{BEM}$) with the same parameters ($R_0=50$~nm, $\xi=10$~nm, $d=100$~nm and acceleration voltage 30~kV); we further considered $R_\mathrm{c}=35$~nm. (a1) The absolute value of the analytical $w$ for the first mode, and (a2) the relative comparison with the MNPBEM calculation. Circles (a2, b2, and c2) represent the area where 95\% of a beam intensity lies for Laguerre-Gaussian beams ($\BW=10$~nm, energy 30~keV) with a different initial OAM. (a3) The phase of $w$ for the first mode, and (a4) the difference between the phase calculated by MNPBEM. (b1--b4) Same as (a1--a4) for the second mode, and (c1--c4) same as (a1--a4) for the third mode. In (b2) and (c2), the colourmap is saturated at -100\%. These divergences in the comparison sub-figures are caused by dividing two numbers close to zero [see the areas where the $|w_\mathrm{a}|$ is close to zero in (b1) and (c1)].
    }
    \label{fig:SI_Fig4}
\end{figure}

\begin{figure}
    \centering
    \includegraphics[width=1\linewidth]{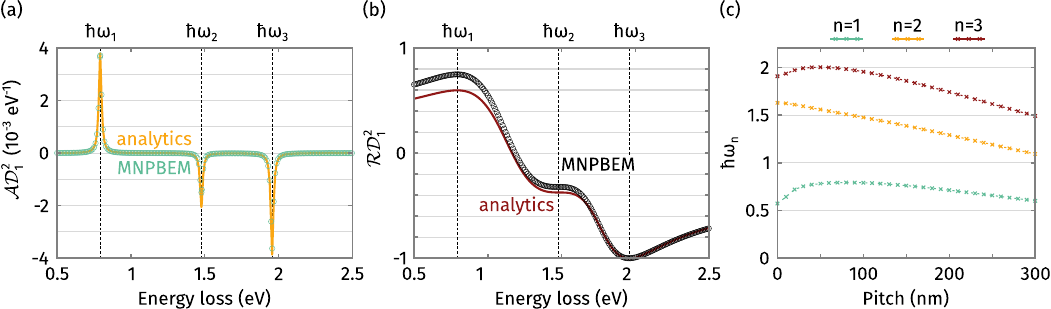}
    \caption{(a) Comparison of the absolute dichroism $\mathcal{AD}_{1}^{2}$ (a), and relative dichroism $\mathcal{RD}_{1}^{2}$ (b) calculated numerically by MNPBEM and using approximated model with infinitesimally thin silver single-twist helix with a charged distribution described by cosine dependence with parameters $R_0=50$~nm, $\xi=10$~nm, $d=100$~nm, acceleration voltage 30~kV, and beam waist $\BW$=10~nm.  (c) The mode energies for the first three plasmonic modes are dependent on the helix pitch. Data were calculated by MNPBEM and used in analytical calculations.}
    \label{fig:SI_Fig7}
\end{figure}


\begin{figure}
    \centering
    \includegraphics[width=1\linewidth]{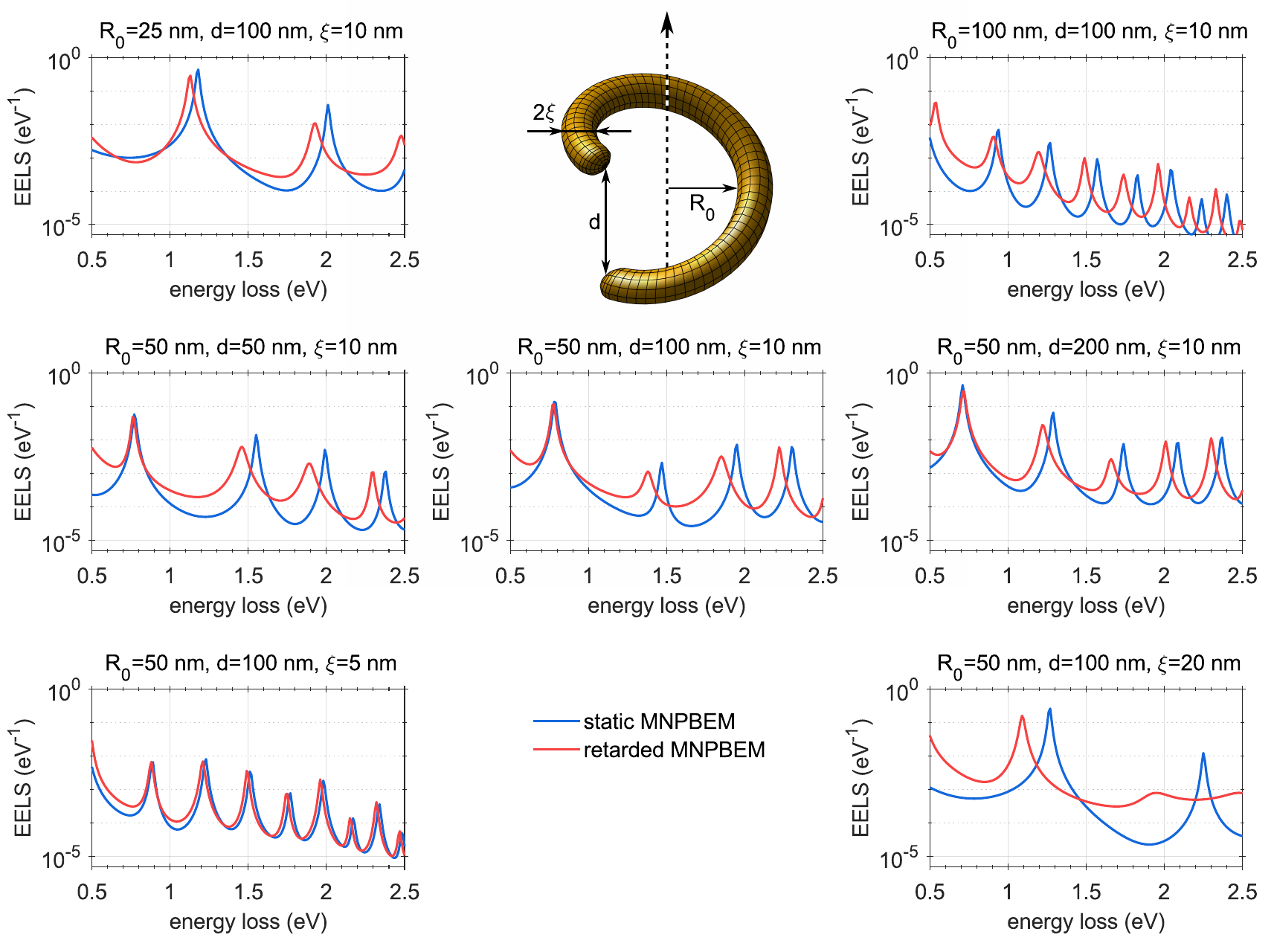}
    \caption{Electron energy-loss spectra for helix nanoparticles with various geometrical parameters calculated using the retarded (red) and static (blue) MNPBEM. Acceleration voltage was fixed at 30~kV, and the lateral position of the electron beam was set to $\Rb_{\mathrm{b}} = [ 0.1\,R_{0}, 0.1\,R_{0} ] $. The plotted spectra show that in terms of mode energies, the static BEM yields similar values to those calculated by the retarded variant, with relative errors usually within $10\%$. The difference in mode energies and peak amplitudes becomes more pronounced for thicker helices (the last plot in the third row), but generally speaking, the static BEM provides---at least within the parametric space explored here---a good estimate of the helix response and its dichroic interaction with vortex electron beams.}
    \label{fig:SI_Fig9}
\end{figure}



\end{widetext}

\end{document}